\documentclass[useAmain sequence,usenatbib]{mn2e}

 \usepackage{graphicx}
 \newcommand{\href}[2]{#2}
 \newcommand{\url}[1]{\texttt{#1}}

\usepackage{subfigure}
\usepackage{float}
\usepackage{placeins}
\usepackage{bm}
\usepackage{longtable}
\usepackage{pdflscape}

\DeclareMathAlphabet{\mathsc}{OT1}{cmr}{m}{sc}
\def\testbx{bx}
\DeclareRobustCommand{\ion}[2]{
\relax\ifmmode
\ifx\testbx\f@series
{\mathbf{#1\,\mathsc{#2}}}\else
{\mathrm{#1\,\mathsc{#2}}}\fi
\else\textup{#1\,{\mdseries\textsc{#2}}}
\fi}
\newcommand{\Hi} {\ion{H}{i}}

\newcommand{\ha} {\mbox{H$\alpha$}}
\newcommand{\hb} {\mbox{H$\beta$}}
\newcommand{\hg} {\mbox{H$\gamma$}}

\newcommand{\Hei} {\ion{He}{i}}
\newcommand{\Heii} {\ion{He}{ii}}

\newcommand{\Ni} {\ion{N}{i}}

\newcommand{\Oia} {[\ion{O}{i}]}
\newcommand{\Oi} {\ion{O}{i}}

\newcommand{\Nai} {\ion{Na}{i}}

\newcommand{\Siii} {\ion{Si}{ii}}

\newcommand{\Caiia} {[\ion{Ca}{ii}]}
\newcommand{\Caii} {\ion{Ca}{ii}}
\newcommand{\Scii} {\ion{Sc}{ii}}
\newcommand{\Tiii} {\ion{Ti}{ii}}

\newcommand{\Feiia} {[\ion{Fe}{ii}]}
\newcommand{\Feii} {\ion{Fe}{ii}}

\newcommand{\Baii} {\ion{Ba}{ii}}

\newcommand{\sn}{SN 2013ab}
\newcommand{\host}{NGC 5669}

\newcommand{\synow}{\textsc{synow}}

\newcommand{\iraf}{\texttt{IRAF}}
\newcommand{\daophot}{\textsc{daophot}}

\newcommand{\cmfgen}{\textsc{cmfgen}}

\newcommand{\epm}{\textsc{epm}}
\newcommand{\ubvri}{\textit{UBVRI}}
\newcommand{\swift}{\textit{Swift}}

\newcommand{\ebv}{\mbox{$E(B-V)$}}

\newcommand{\msun}{\mbox{M$_{\odot}$}}

\newcommand{\rsun}{\mbox{R$_{\odot}$}}
\newcommand{\kms}{\mbox{$\rm{\,km\,s^{-1}}$}}
\newcommand{\ergs}{\mbox{$\rm{\,erg\,s^{-1}}$}}

\newcommand{\nickel}{\mbox{$^{56}$Ni}}
\newcommand{\cobalt}{\mbox{$^{56}$Co}}
\newcommand{\iron}{\mbox{$^{56}$Fe}}
\newcommand{\mum}{\mbox{$\mu{\rm m}$}}

\begin{document}

\title[Type IIP Supernova 2013ab]
{\sn\ : A normal type IIP supernova in \host}
\author[Bose et al.]
{Subhash Bose$^{1,2}$\thanks{e-mail: email@subhashbose.com, bose@aries.res.in},
 Stefano Valenti$^{3,4}$, Kuntal Misra$^1$, Maria Letizia Pumo$^{5,6,7}$,
 \newauthor
 Luca Zampieri$^5$, David Sand$^8$, Brijesh Kumar$^1$, Andrea Pastorello$^5$,
 \newauthor
 Firoza Sutaria$^9$, Thomas J. Maccarone$^8$, Brajesh Kumar$^{1,10}$, M. L. Graham$^{11}$,
 \newauthor
 D. Andrew Howell$^{3,4}$, Paolo Ochner$^5$, H. C. Chandola$^2$, Shashi B. Pandey$^1$
\\
 $^1$Aryabhatta Research Institute of Observational Sciences, Manora
    Peak, Nainital - 263002, India.\\
 $^2$Centre of Advance Study, Department of Physics, Kumaun University, Nainital - 263001, India.\\
 $^3$Las Cumbres Observatory Global Telescope Network, 6740 Cortona Dr., Suite 102, Goleta, CA 93117, USA\\
 $^4$Department of Physics, University of California, Santa Barbara, Broida Hall, Mail Code 9530, Santa Barbara, CA 93106-9530, USA\\
 $ ^5 $INAF, Osservatorio Astronomico di Padova, 35122 Padova, Italy\\
 $^6$Universit\`{a} di Catania, Dip. di Fisica e Astronomia (Sez. astrofisica),
  via S. Sofia 78, 95123 Catania, Italy\\
 $^7$INAF-Osservatorio Astronomico di Palermo, Piazza del Parlamento 1, 90134 Palermo, Italy\\
 $^8$Physics Department, Texas Tech University, Lubbock, TX 79409, USA\\
 $^9$Indian Institute of Astrophysics, Block-II, Koramangala, Bangalore - 560034, India.\\
 $^{10}$Institut d'Astrophysique et de G\'{e}ophysique, Universit\'{e} de
 Li\`{e}ge, All\'{e}e du 6 Ao\^{u}t 17, B\^{a}t B5c, 4000 Li\`{e}ge, Belgium \\
 $^{11}$Department of Astronomy, University of California, Berkeley, CA 94720-3411 USA
}

\date{Accepted.....; Received .....}

\pagerange{\pageref{firstpage}--\pageref{lastpage}} \pubyear{}

\maketitle

\label{firstpage}

\begin{abstract}
 We present densely-sampled ultraviolet/optical photometric and low-resolution
 optical spectroscopic observations of the type IIP supernova 2013ab in the nearby ($\sim$24 Mpc) galaxy \host,  from 2 to 190d after explosion.
 Continuous photometric observations, with the cadence of typically a day to one week, were acquired with the 1-2m class telescopes in the LCOGT network, ARIES telescopes in India and various other telescopes around the globe.
 The light curve and spectra suggest that the SN is a normal type IIP event with a plateau duration of $ \sim80 $ days with mid plateau absolute visual magnitude of -16.7, although with a steeper decline during the plateau (0.92 mag 100 d$ ^{-1} $ in $ V $ band)
 relative to other archetypal SNe of similar brightness. The velocity profile of \sn\ shows striking resemblance with those of SNe 1999em and 2012aw.
 Following the \cite{2011ApJ...728...63R} prescription, the initial temperature evolution of the SN emission
 allows us to estimate the progenitor radius to be $ \sim $ 800 \rsun, indicating that the SN originated from a red supergiant star.
 The distance to the SN host galaxy is estimated to be  24.3 Mpc from expanding photosphere method (\epm).
 From our observations, we estimate that 0.064 \msun\ of \nickel\ was synthesized in the explosion.
 General relativistic, radiation hydrodynamical modeling of the SN infers an explosion energy of $ 0.35\times10^{51} $~erg, a progenitor mass (at the time of explosion) of $ \sim9 $~\msun\ and an initial radius of $ \sim600 $~\rsun.

\end{abstract}


\section{Introduction} \label{sec:intro}

Type IIP supernovae (SNe) are a sub-class of Core-Collapse SNe (CCSNe) whose progenitors had retained substantial amount of hydrogen
before they exploded as SNe. These SNe exhibit nearly constant brightness in their light curve for a few months after explosion, known as plateau phase. This is explained as a combined effect of expanding ejecta  and receding recombination layer of hydrogen due to adiabatic cooling of the envelope \citep[e.g.][]{2009ApJ...703.2205K}. After reaching the end of plateau phase, a type IIP SN light curve shows a very steep decline and then finally settles on to a relatively slowly declining phase labelled as the radioactive tail. This stage, also called the nebular phase, is powered by the radiation originating from the radioactive decay of \cobalt\ to \iron\ \citep{1980ApJ...237..541A} which in turn depends upon the amount \nickel\ synthesized in the explosion.

Type IIP SNe are also proven candidates for distance estimation at extragalactic scales. \cite{2001PhDT.......173H} explored the potential of these SNe as standardizable candles. The expanding photosphere method (\epm) \citep{1974ApJ...193...27K}  for estimating distances has   also been applied extensively. Significant contribution has been made to improve and implement \epm\ by several authors, viz. \cite{2001ApJ...558..615H,2005A&A...439..671D,2009ApJ...696.1176J,2014ApJ...782...98B}. \epm\ requires an extensive set of photometric and spectroscopic data of SNe.

% tab:host
%____________________________________________________________________________

  \begin{table}
  \caption{Properties of the \sn\ and its host galaxy \host.}
  \label{tab:host} 
  %\centering

  \begin{tabular}{llc} \hline \hline
     \noalign{\smallskip}
      Parameters& Value& Ref.$^{a}$\\ 
     \noalign{\smallskip} \hline
    
     \noalign{\smallskip}
     \multicolumn{3}{l}{\bf \host:}\\
     Type& Sb& 2\\
     RA (J2000)& $\alpha = 14^{\rm h} 32^{\rm m} 43\fs8$& 2\\
     DEC (J2000)& $\delta = 09\degr 53\arcmin 28\farcs8$& 2\\
     Abs. Magnitude& $M_{B}=-19.28$ mag& 2\\
     \\
     Distance& $D=24.0\pm0.9$ Mpc& 1 \\
     %?Scale& $1\arcsec \sim 48$ pc, ~~$1\arcmin \sim 2.9$ kpc&\\
     Distance modulus& $\mu = 31.90\pm0.08$ mag&\\
     \\
     %?Apparent radius&${r_{\mathrm{25}}}=3\farcm6\,(\sim 10.5\,\mathrm{kpc})$& 1\\
     %?Inclination angle&$\Theta_{\rm inc}= 54.6^\circ$&1\\
     %?Position angle& $\Theta_{\rm maj}=9.9^\circ$&1\\
     Heliocentric Velocity& $cz_{\rm helio}=1374\pm2 \kms$&2\\
     \\
     \multicolumn{3}{l}{\bf \sn:}\\
     RA (J2000)& $\alpha = 14^{\rm h} 32^{\rm m} 44\fs49$& 3\\
     DEC (J2000)& $\delta = 09\degr 53\arcmin 12\farcs3$& \\
     \\
     Galactocentric Location& 7\arcsec.5 E, 18\arcsec.1 S&  \\
     %?Deprojected radius&  $r_{\rm SN} = 139\farcs1$ ($\sim$ 6.75 kpc)& \\
     \\
     Time of explosion & $t_{\rm 0} =$ 16.5 February 2013 (UT)& 1\\ 
                    & (JD $ 2456340.0\pm1.0 $) day& \\ 
     Total reddening & \ebv\,= $0.044\pm0.066$ mag& 1\\
   
     \noalign{\smallskip}
     \hline
  \end{tabular}
  \newline (1) This paper;
        (2) HyperLEDA - http://leda.univ-lyon1.fr \citep{2014A&A...570A..13M};
  		(3) \citet{2013ATel.4823....1Z}
                  %(2) \citet{2012ApJ...756..131V}
  \end{table}

According to our current understanding of stellar evolution, type IIP SNe originate from red supergiant (RSG) stars having initial masses of 9-25\msun, with an upper mass limit of $ \sim 32$~\msun, for solar metallicity
stars \citep[e.g.][]{2003ApJ...591..288H}.
Recent studies also identified Super-AGB stars as potential progenitors of some SNe IIP \citep[see e.g.][and references therein]{2009ApJ...705L.138P}.
However there is a limited number of nearby SNe where high resolution pre-SN $ HST $ images are available. The estimated range of masses for these progenitors from direct observations lies within 9-17 \msun\ \citep{2009MNRAS.395.1409S}. On the other hand, hydrodynamical modeling of a handful of well-studied SN light curves suggests that their progenitor masses are within 15-25 \msun\ \citep{2009A&A...506..829U,2011ApJ...729...61B}.

The cosmological importance of type IIP SNe and the ambiguity in the present understanding of their evolution and physical mechanisms are key motivations to study individual events with a range of properties.
One of the best observed recent type IIP event is SN 2013ab.

\begin{figure}
\centering
\includegraphics[width=\linewidth]{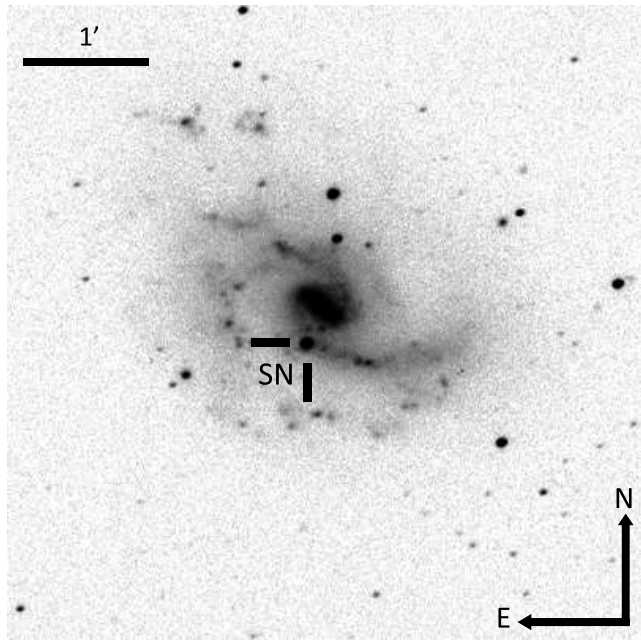}
\caption{\sn\ in \host. The $V$-band image taken from 104-cm ST covering a subsection of
         about 5\arcmin.1$\times$5\arcmin.1 is shown.}
\label{fig:snid}
\end{figure}

 \sn\ was discovered on February 17.5 UTC, 2013 by \cite{2013CBET.3422....1B} in the
 galaxy \host\ ($\sim$ 25 Mpc) (see Fig.~\ref{fig:snid}) at $R \sim 17.6$ mag. The last non-detection
 was reported on February 15 \citep{2013ATel.4823....1Z} to a limiting magnitude $R\sim18.5$.
 We therefore adopt February 16.5, 2013 (JD=2456340.0 $\pm$ 1.0 days) as the time of explosion (0d phase)
 throughout the paper. Some basic parameters of \sn\ and its host galaxy  are listed in Table~\ref{tab:host}.

 In this work, we present results from optical photometric (\ubvri\ and \textit{gri})
 follow-up observations of \sn\ at 136 phases (from 3 to 190d), \swift~UVOT observations at 25 epochs (from 4 to 103d) and low-resolution optical spectroscopic
 observations at 25 phases (from 2 to 184d). The paper is organized as follows.
 The \S\ref{sec:obs} provides details of the photometric and spectroscopic
 observations. Determination of reddening and extinction is described in \S\ref{sec:ext}.
 In \S\ref{sec:lc}, we study the light and colour curves, derive bolometric light curves whose tail luminosities are used to estimate the $^{56}$Ni mass. In \S\ref{sec:sp}, we study  the spectral evolution, present
 \synow\ modelling and derive line velocities.
 Application of \epm\ and estimate of distance to SN is described in \S\ref{sec:epm}.
 Hydrodynamical modeling to estimate physical parameters is described in \S\ref{sec:char}. A brief summary of the work is  given in \S\ref{sec:sum}.

\section{Observations} \label{sec:obs}

 Broadband photometric data have been collected in Johnson $ BV $ and Sloan $ gri $ systems using the Las Cumbres Observatory Global Telescope (LCOGT) network, description of instrument and telescopes are presented in \cite{2013PASP..125.1031B}.
 We have also used ARIES 104-cm Sampurnanand Telescope (ST) and the 130-cm Devasthal
 Fast Optical Telescope (DFOT) to acquire broadband data in Johnson-Cousin \ubvri\ filters. The instrument details are presented in \cite{2013MNRAS.433.1871B} and \cite{2012ASInC...4..173S}.
 \swift~UVOT has also observed \sn\ in ultra-violet (UV) and optical broadbands. A detailed description of data reduction and derivation of  photometric magnitudes are given in Appendix \ref{app:photometry}.

Low resolution spectroscopic observations have been carried out at 25 phases from 2 to 184d after explosion: 12 epochs of data were collected using Floyds spectrograph on Faulkes Telescope North (FTN), 6 epochs on Faulkes Telescope South (FTS), 5 epochs using HFOSC on Himalayan Chandra Telescope (HCT) and 2 epochs using the B\&C spectrograph mounted on Galileo Telescope in Asiago. The data reduction process is given in Appendix~\ref{app:spectroscopy} and the journal of spectroscopic observations in Table~\ref{tab:speclog}.

\section{Extinction and Distance} \label{sec:ext}

 In order to derive intrinsic properties of the explosion, the line-of-sight reddening of \sn\ due to interstellar dust
 in both the Milky Way and the host galaxy should be known
 accurately. Using all-sky dust-extinction map of \cite{2011ApJ...737..103S}\footnote{http://irsa.ipac.caltech.edu/applications/DUST/}, we adopt
 the following value of Galactic reddening: $\ebv_{\rm MW}$ = $0.0234\pm0.0002$ mag.

 \begin{figure}
 \centering
  \includegraphics[width=8.4cm]{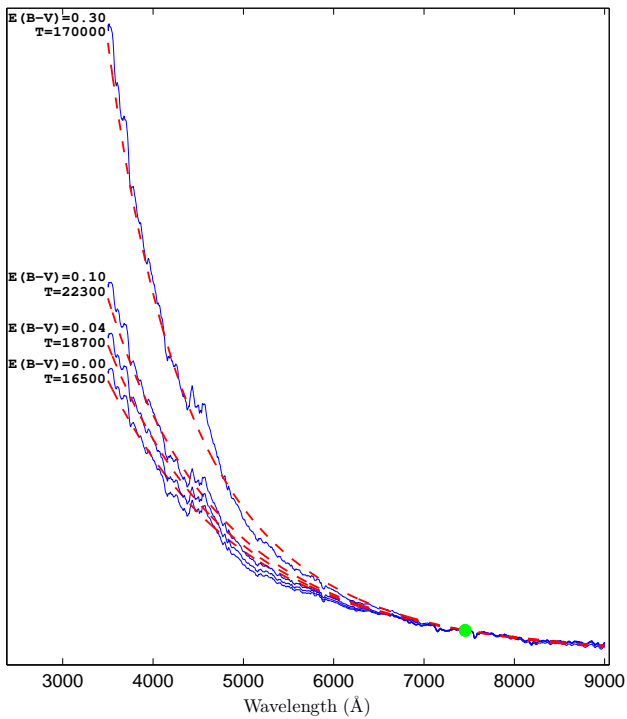}
 \caption{The spectral energy distribution of \sn\, at 2.2d is compared with a blackbody function (dotted line). The fluxes are normalized relative to an arbitrary line-free region in the spectra marked with a green filled circle. Temperature units are kelvin (K) and \ebv\ in magnitude.}
 \label{fig:ebv.sed}
 \end{figure}

One of the widely adopted techniques for reddening estimate is using the narrow \Nai~D interstellar absorption dips. The equivalent width (EW) of \Nai\,D absorption feature is found to be correlated with the reddening \ebv\, estimated from the tail of SN Ia colour curves \citep{1990A&A...237...79B,2003fthp.conf..200T}.
However in our low resolution set of spectra, we did not identify any \Nai~D absorption
feature indicating that extinction due to the host galaxy is very low and of the order of Galactic reddening.
To further constrain \ebv\ due to the host galaxy, we implement ``colour-method" \citep{2010ApJ...715..833O}, which assumes that at the end of plateau phase the intrinsic $ (V-I) $ colour is constant. Thus, it is possible to obtain the host colour excess from observed $ (V-I) $ colour and relate it to the visual extinction. The relation found by \cite{2010ApJ...715..833O}
is :
\begin{eqnarray}
A_V(V-I)&=&2.518[(V-I)-0.656] \\
\sigma_{(A_V)}&=&2.518\sqrt{\sigma^2_{(V-I)}+0.053^2+0.059^2}
\end{eqnarray}
Using the mean $ (V-I) $ colour within 78-82 d (corresponds to the end of plateau phase), and correcting it for Galactic reddening, we obtain $ A_{V_{host}}=0.0624\pm0.2060 $ mag which corresponds to \ebv$ _{host}=0.0201\pm0.0664 $ mag assuming total-to-selective extinction $R_{\rm V}=3.1$.
Hereafter, we adopt a total \ebv~$= 0.044\pm0.066 $ mag along the line of sight to \sn\ providing a total $ A_V= 0.14\pm0.21$ mag.

 To seek further justification of the derived \ebv\, we look into the earliest spectra at 2.2 d. At early phases the spectral energy distribution (SED) can be well approximated as a blackbody. Hence we de-redden the spectra with different values of \ebv\, and estimate corresponding blackbody temperatures (see Fig.~\ref{fig:ebv.sed}). For $ \ebv=0.30 $ mag we obtain an unphysically high temperature of 170 kK. The theoretical modelling of \cite{2006A&A...447..691D,2011ApJ...729...61B} indicate that for a 2.2d old type IIP SN, the temperature must be around 25-30 kK.
 Our blackbody fit to the spectra with $ \ebv=0.10 $ mag results in a temperature estimate of 22.3 kK. This is consistent  with the values predicted by theoretical modeling.
 Also we estimate a temperature of 18.7 kK corresponding to our adopted $ \ebv=0.044 $ mag. This analysis provides an approximate upper limit of $ \ebv=0.10 $ mag which is consistent with the adopted reddening value determined using colour-method.
 \\

A number of distance estimates to \host\, using the Tully-Fisher method are available in the literature with a wide variation in values ranging from 18 to 32 Mpc.
Hence to seek for a
reliable estimate, we applied \epm\ to the SN and derived a distance of $ 24.26\pm0.98 $ Mpc. The detailed \epm\ analysis will be discussed in \S\ref{sec:epm}. We adopt the distance to host galaxy to be $ 24.0\pm0.9 $ Mpc which is the weighted mean of EPM and two other recent Tully-Fisher estimates from the literature, viz. \cite{2007A&A...465...71T} ($ 25.23\pm4.65 $ Mpc) and \cite{2009AJ....138..323T} ($ 19.67\pm3.35 $ Mpc), assuming $ H_0=73 $ km~s$ ^{-1} $Mpc$ ^{-1}$, $ \Omega_m=0.27 $, $ \Omega_\Lambda=0.73 $.

\section{Optical light curve} \label{sec:lc}

 \subsection{Apparent magnitude light curves} \label{sec:lc.app}

\begin{figure}
\centering
\hspace*{-4.0mm}
\includegraphics[width=1.05\linewidth]{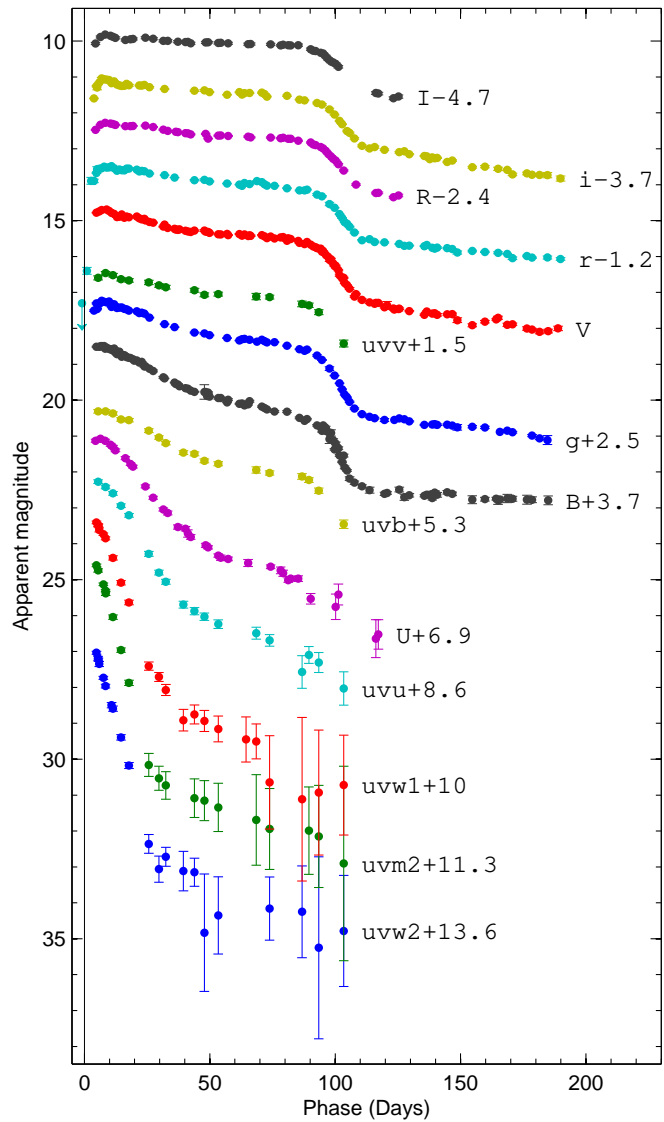}
\caption{The photometric light curve of \sn\ in Johnson-Cousins \ubvri, SDSS \textit{gri} and \swift~UVOT bands. The light curves are shifted arbitrarily for clarity. The large errors in late UV data points are due to faint SN flux extracted after subtracting host background.}
\label{fig:lc.app}
\end{figure}

 Photometric measurements in Johnson-Cousins \ubvri\ and SDSS \textit{gri} are available at 136 phases from 1 to 189d after the explosion, with a stringent non-detection at -1d. Additional 25 epoch observations are from \swift~UVOT  in all six UVOT filters. The resulting light curves are shown in Fig.~\ref{fig:lc.app} and data is tabulated in Table~\ref{app.data}.

 The early light curve initially shows a sharp rise in \textit{r} band, which is also visible in all other optical bands as well, but only during first few phases. Then the light curve declines slowly until the end of plateau-phase. Since 95d, a steep decline to the radioactive nebular tail follows. After that, since $\sim$ 113d, the light curve settles on to this relatively slower declining phase. Observations in the UVOT bands do not show any initial rise in the light curve, although observations started with the same delay as the optical bands (+4d). The early peak is found to occur at 6.4, 7.2, 7.8, 8.3, 8.3, 7.8, 8.4 and 7.8d in \textit{UBVRIgri} bands respectively, with uncertainties of about 1d. This is consistent with most fast-rising SNe \citep[e.g SN 2005cs,][]{2009MNRAS.394.2266P}, and is significantly different from some SNe which exhibit a slow-rising early phase light curve (see e.g. the delayed \textit{V}-band maximum attained at 16d in SN 2006bp \citep{2007ApJ...666.1093Q}; 13d in SN 2009bw \citep{2012MNRAS.422.1122I}; and 15d in SN 2012aw \citep{2013MNRAS.433.1871B};).

 The decline rates after the initial maximum to the plateau-end in \textit{UBVRIgri} are
 7.60, 2.72, 0.92, 0.59, 0.30, 1.68, 0.77, 0.51 mag 100 d$ ^{-1} $ respectively. This is steeper than the values reported for SN 1999em
 \citep{2002PASP..114...35L}, SN 1999gi \citep{2002AJ....124.2490L} and SN 2012aw \citep{2013MNRAS.433.1871B}. For example in the $ UBV $-bands, SN 2012aw experienced a decline rate of 5.60, 1.74, 0.55 mag 100 d$ ^{-1} $. However, the decline rate of SN 2013ab is similar to that of SN 2004et \citep[2.2 mag 100 d$ ^{-1} $ in $ B $ band;][]{2006MNRAS.372.1315S}.
 During the nebular phase, the decline rate (mag 100 d$ ^{-1} $) of the light curves
 are estimated to be 0.36, 0.97, 0.76, 0.66 and 1.16 for \textit{BVgri} respectively.

 \sn\ is luminous in the UVOT $UV$ bands at early phases, but it declines steeply at a rate of 0.169, 0.236 and 0.257 mag d$ ^{-1} $ in \textit{uvw1, uvw2 and uvm2} bands respectively. After 30d, the light curves settle on a slow-declining plateau until $\sim$ 103d corresponding  also with the end of \swift~UVOT observations. SN 2012aw is the only known SN which shows a UV plateau \citep{2013ApJ...764L..13B} similar to that observed in \sn. Although, a UV plateau is expected, not many type II SNe have been observed so far at these wavelengths until relatively late phases. This is possibly because of their low apparent brightness due to large distance and extinction, making them unsuitable for UVOT detections. However, these limitations did not hinder \sn\,
 making this event as one of the best observed SNe IIP in the UV domain.

 \subsection{Absolute magnitude and colour evolution} \label{sec:lc.abs}

\begin{figure}
\centering
\includegraphics[width=8.5cm]{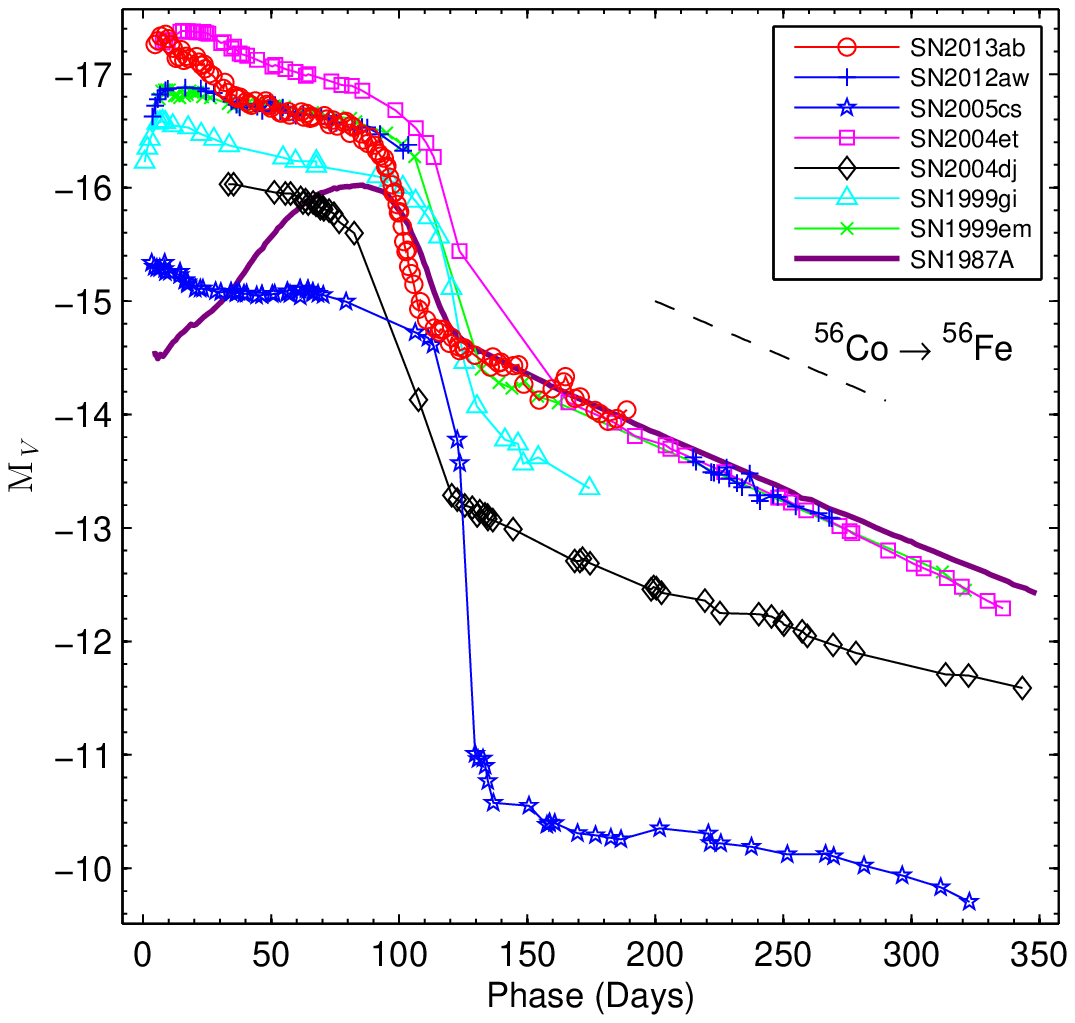}
\caption{Comparison of the $ V $-band absolute light curve of {\sn} with those of other type IIP SNe. The exponential decline
         of the radioactive decay law is indicated with a dashed line. The time of explosion in JD-2400000,
         distance in Mpc, reddening \ebv\, in mag  and
  the reference for apparent V-band magnitude, respectively, are :
  SN 1999em -- 51475.6, 11.7, 0.10, \citet{2002PASP..114...35L,2003MNRAS.338..939E};
  SN 2004et -- 53270.5, 5.4, 0.41, \citet{2006MNRAS.372.1315S};
  SN 2005cs -- 53549.0, 7.8, 0.11, \citet{2009MNRAS.394.2266P};
  SN 2004dj -- 53187.0, 3.5, 0.07, \citet{2008PZ.....28....8T};
  SN 1987A  -- 46849.8, 0.05, 0.16,  \citet{1990AJ.....99.1146H};
  SN 1999gi -- 51522.3, 13.0, 0.21, \citet{2002AJ....124.2490L};
  SN 2012aw -- 56002.6, 9.9, 0.07, \citet{2013MNRAS.433.1871B}.}
\label{fig:lc.abs}
\end{figure}

The $ V $-band absolute light curve of \sn\ is shown in Fig.~\ref{fig:lc.abs}, and is compared with those of other well-studied type IIP SNe. All data are corrected for their corresponding distances and extinction values. \sn\ is compared with the normal SNe 1999em, 1999gi, 2004dj, 2004et and 2012aw; the subluminous  SN 2005cs and the photometrically peculiar SN 1987A. The comparison shows that the V-band mid-to-late plateau absolute magnitude of \sn\ is very similar to those of SNe 1999em and 2012aw. However the plateau light curve decay rate, especially during the early-plateau (from 10 to 50d) phase is significantly larger than those of SNe 1999em and 2012aw (by 1.58 and 2.61 times respectively), although later on (during late-plateau phase) the slopes are somewhat similar.
The decay rate  of the early-plateau light curve is as high as 1.58 mag 100 d$ ^{-1}$, which is significantly higher than that of the late-plateau light curve (0.49 mag 100 d$ ^{-1} $).
The nebular-phase light curve evolution follows a decay rate of 0.97 mag 100 d$ ^{-1} $ which is similar to those of other SNe in our comparison sample. This is consistent with the expected decay rate of the $ ^{56} $Co to $ ^{56} $Fe (0.98 mag 100 d$ ^{-1} $).
The mid-plateau absolute magnitude is M$^{p}_{V}=-16.7$ mag, categorizing \sn\ as a normal type IIP event \citep{1994A&A...282..731P}. This magnitude make \sn\ significantly brighter than sub-luminous class of events like SN 2005cs \citep[M$^{p}_{V}$ $\sim$ $-15$ mag;][]{2009MNRAS.394.2266P}.
Another noticeable difference with the compared SNe is the relatively shorter plateau duration, $ \sim $78d in \sn, in contrast to $ \sim $90d and 92d for SNe 1999em and 1999gi respectively.
\cite{2014ApJ...786...67A} found an anti-correlation between the slope of early plateau and full-plateau duration for type II SNe, which is consistent with the faster decline but shorter plateau duration of \sn. However, it fits somewhat at the lower end of the scatter relation.

 \begin{figure}
 \centering
 \includegraphics[width=1.02\linewidth]{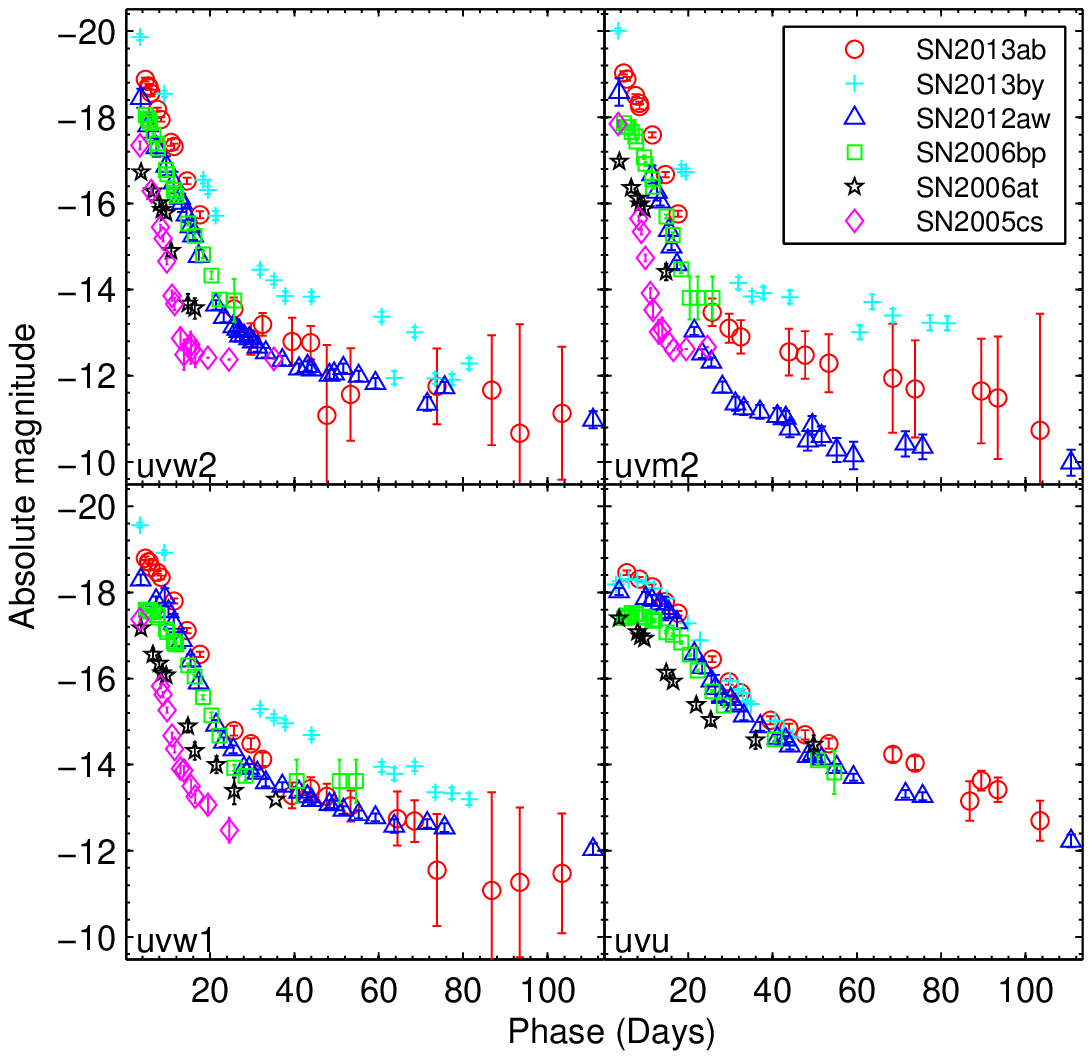}
 \caption{\swift~UVOT UV absolute light curves of \sn, are compared with other well observed IIP SNe from UVOT. For the compared SNe, references for UVOT data, extinction and distance are: SN 2005cs -- \citet{2009AJ....137.4517B,2009MNRAS.394.2266P}, SN 2006at -- \citet{2009AJ....137.4517B}; Distance 65 Mpc; $ \ebv=0.031 $ mag \citep[only Galactic reddening][]{2011ApJ...737..103S}, SN 2006bp -- \citet{2008ApJ...675..644D}, SN 2012aw -- \citet{2013ApJ...764L..13B,2013MNRAS.433.1871B}, SN 2013by -- \citet{2015MNRAS.448.2608V}.}
 \label{fig:uv.abs}
 \end{figure}

\swift~UVOT absolute light curves (in \textit{uvu, uvw1, uvm2} and \textit{uvw2} bands) of \sn\ are shown in Fig.~\ref{fig:uv.abs} and are compared with other well observed IIP SNe. Distance and extinction has been corrected for each of the events. For SN 2006at, extinction is not known, hence a minimal reddening has been adopted accounting only for Milky Way extinction. \sn\ is on the brighter end among the compared events. Most  SNe were not detected in UV after about a month, this is primarily due to large distances and  extinction values. Both of these factors being not a major issue in \sn, which has been in fact observed for more than 100d. SNe 2012aw and 2013by are comparable with \sn\ in terms of data coverage and clear detection of a plateau in the UV domain. The plateau is evident in all UV bands after 30d and follows a similar trend as that observed in SN 2012aw.

\begin{figure}
 \centering
 \includegraphics[width=8.5cm]{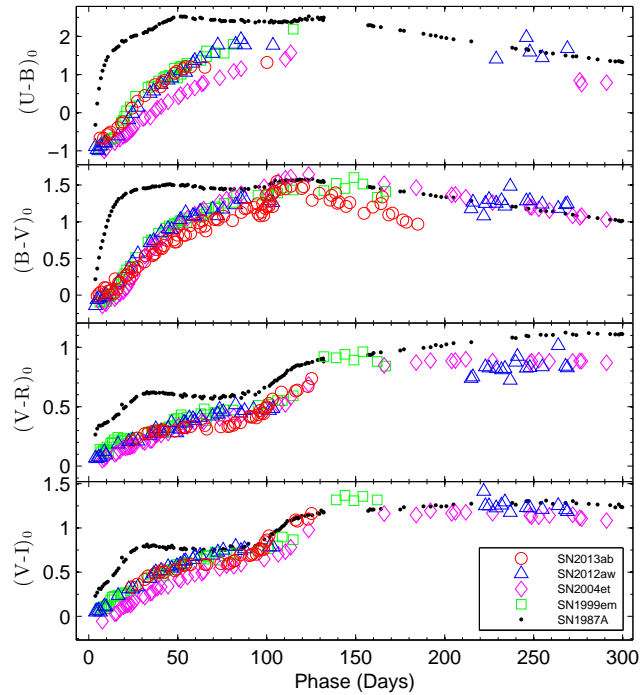}
 \caption{The evolution of intrinsic colours of {\sn} is compared with those of other well-studied
          type IIP SNe 1987A, 1999em, 2004et and 2012aw. Reference for data are the same as in
          Fig.~\ref{fig:lc.abs}.}
 \label{fig:cc.abs}
 \end{figure}

The broadband colour evolution provides important information about the temporal variation of the SN envelope properties. The expansion and cooling behavior of the envelope can be studied from the colour evolution at different phases. The intrinsic colour evolutions (\textit{U-B}, \textit{B-V}, \textit{V-R}
and \textit{V-I}) are shown in Fig.~\ref{fig:cc.abs}. All colours show a rapid evolution towards redder colours until $ \sim $ 50d, due to a rapid cooling of fast expanding ejecta. Thereafter, they evolve relatively slowly until the onset of the nebular-phase. The colour evolution is very similar in other archetypal SNe IIP such as SN 1999em and SN 2012aw.
The $ (B-V) $ colour shows a bluer trend after 120d, when the nebular-phase begins. In this phase, the ejecta become sufficiently optically thin to allow photons from radioactive decay of $ ^{56} $Co to $ ^{56} $Fe to escape.

 \begin{figure}
 \centering
 \includegraphics[width=8.4cm]{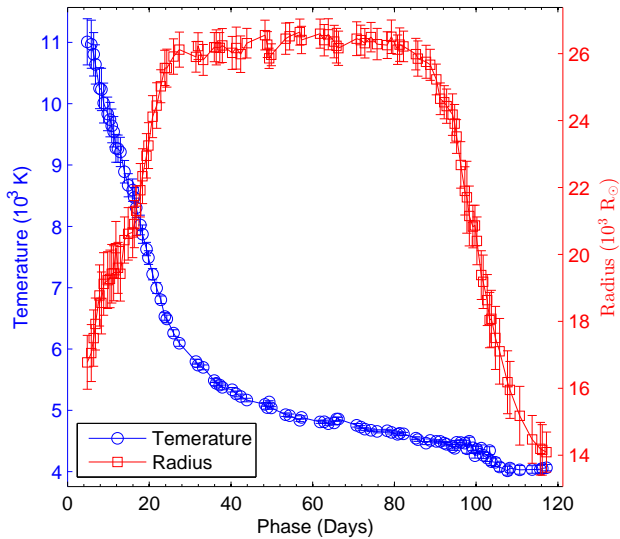}
 \caption{The evolutions of temperature and radius in \sn, as derived from blackbody fits to
          the observed fluxes in the optical range (0.36-0.81 $\mum$).}
 \label{fig:lc.rad}
 \end{figure}

 To have an idea of the temporal evolution of temperature and photospheric radius, we fit blackbody functions to the broadband optical fluxes (after correcting for total line-of-sight extinction). The blackbody radii are further corrected by dilution factors \citep[using the prescription of][]{2005A&A...439..671D} to estimate photospheric radii (where optical depth is $ \tau=2/3 $) rather than thermalization radii. The plot with the photospheric temperature and radius evolutions is shown in Fig.~\ref{fig:lc.rad}. The temperature drops very rapidly from 5 to 25d due to adiabatic cooling of the rapidly expanding envelope. Thereafter, the decline flattens as the SN progressively enters the nebular phase. The photospheric radius increases rapidly as the SN expands and thereafter the radius remains almost constant until around 85d, which marks the end of plateau phase. This apparent contradiction to the expansion is due to the fact, that with the fall of temperature the ionized hydrogen starts to recombine by depleting the free electrons thereby
 optically thinning the outer ejecta. This results to a receding photospheric layer on top of the expanding envelope, ultimately leading to an unchanged photospheric radius. After 85d the radius falls off very rapidly with the end of the hydrogen recombination. During this phase, the ejecta become cooler and almost optically
 thin leaving behind no free electrons. Moreover, the dilution factor corrections are no longer applicable as radiation does not have thermal origin.

 \subsection{Bolometric light curve} \label{sec:lc.bol}

 \begin{figure}
 \centering
 \includegraphics[width=8.5cm]{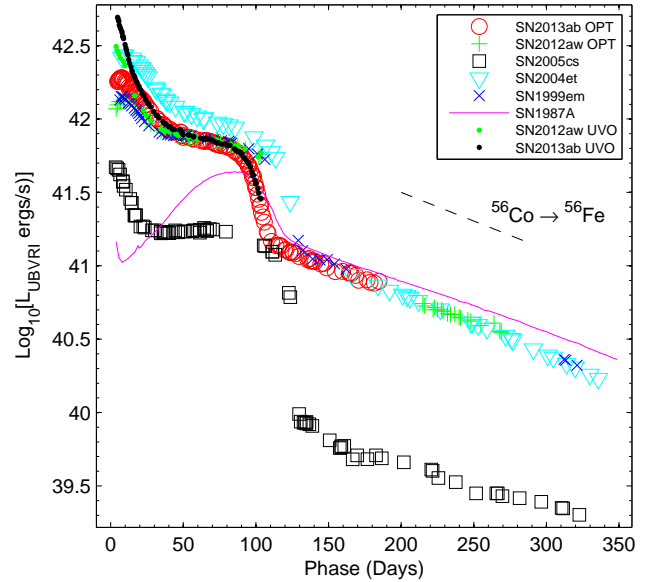}
 \caption{The \ubvri\ pseudo-bolometric light curve of {\sn} is compared with those of other well studied SNe. Light curves with added UVOT UV contributions are also shown for SN 2013ab and SN 2012aw (labeled as UVO).
          The adopted distances, reddening and time of explosion values are same as in Fig.~\ref{fig:lc.abs}. The
          exponential decline of the radioactive \cobalt\ decay law is shown with a dashed line.}
 \label{fig:lc.bol}
 \end{figure}

Pseudo-bolometric luminosities have been computed at all phases adopting the same method as described in \cite{2013MNRAS.433.1871B}, i.e. by constructing a SED from the extinction corrected photometric fluxes which are semi-deconvolved from broadband filter responses. At early phases ($ \le 30$d), when the SN is hot, the bolometric fluxes are dominated by the UV bands. At later phases ($ \ge $ 100d) a major fraction of the bolometric contribution comes from infrared domain. The luminosity is computed within the optical domain (3335 -- 8750\AA), which includes \ubvri\ and \textit{gri} contributions. We have also used UVOT data in the bolometric luminosity computation which covers the wavelength range from the $ uvw2 $ to the $ I $ bands (1606 -- 8750 \AA). The contribution from the UV fluxes yields a significantly higher value of the early luminosity.

In Fig.~\ref{fig:lc.bol}, we plot the optical pseudo-bolometric luminosities of \sn\ along with other well-studied objects, including SNe 1987A, 1999em, 2004et, 2005cs, 2012aw. To have homogeneity in the comparison of bolometric light curves, all luminosities have been computed using the same algorithm and  wavelength range. We also include UV-Optical bolometric light curves of SNe 2013ab and 2012aw for comparison.
The bolometric luminosity declines rapidly by 0.4 dex from 8 to 50d and then goes down further but slowly by 0.1 dex until 85d. It is evident from the comparison that the plateau bolometric luminosity of \sn\ is close to those of SNe 1999em and 2012aw but with a much steeper decline rate of the light curve during the plateau. However, the decline rate and shape of light curve  matches well with that of SN 2004et.
The UV-optical bolometric light curve of \sn\ also shows a sharp decline by 0.8 dex during the first 50d, which is steeper than that observed for SN 2012aw. Thereafter, it declines relatively slowly and coincides with the optical light curve.
The tail bolometric luminosity is similar to those of SNe 1999em, 2004et and 2012aw, and the slope of the tail is nearly identical to that expected for $ ^{56} $Co to $ ^{56} $Fe radioactive decay. Since this powers the tail luminosity, it is directly proportional to the amount of radioactive \nickel\ synthesized during the explosion.

 \subsection {Mass of nickel} \label{sec:lc.nick}

 The radioactive \nickel\ is produced in CCSNe by the explosive nucleosynthesis of Si to O \citep{1980ApJ...237..541A}. Thus the nebular-phase light curve is mainly powered by the radioactive decay of \nickel\ to \cobalt\ and \cobalt\ to \iron, with $ e $-folding time
  of 8.8d and 111.26d respectively emitting $\gamma$-rays and
  positrons. Hence the tail luminosity will be proportional to the amount of synthesized radioactive \nickel.
  The mass of \nickel\ produced by SN 1987A has been determined with a fair degree of accuracy to be $ 0.075\pm0.005 $ \msun\ \citep{1996snih.book.....A}. By comparing the bolometric luminosity of \sn\ with that of SN 1987A at similar phases, we can infer the amount of synthesized \nickel\ in \sn. Although UVOIR bolometric light curve is available for SN 1987A, we preferred to use our \ubvri\ pseudo-bolometric light curve computed using same algorithm that is used for \sn\, to have uniformity in the comparison. We estimated the \ubvri\ bolometric luminosity of \sn\ at 170d, by making a linear fit over 160 to 181d, to be $8.41\pm 0.72\times 10^{40}$\ergs. Likewise, we estimated the luminosity of SN 1987A at similar phases to be $9.93\pm 0.04\times 10^{40}$\ergs. The ratio of \sn\ to SN 1987A is found to be $0.847\pm0.073$, which gives a  $M_{\rm Ni} = 0.064\pm0.006 $ \msun for \sn.

Assuming the $\gamma$-photons emitted from the radioactive decay of \cobalt\ thermalize the ejecta, the \nickel\ mass can be independently estimated from the tail luminosity, as described by \cite{2003ApJ...582..905H}.
 \begin{eqnarray*}
  M_{\rm Ni} = 7.866\times10^{-44} \times L_{t} \exp\left[ \frac{(t_{t}-t_{0})/(1+z)-6.1}{111.26}\right]\msun,
 \end{eqnarray*}
 where $t_{0}$ is the explosion time, 6.1d is the half-life of \nickel\ and 111.26d is the e-folding time of the \cobalt\ decay. We compute the tail luminosity $L_{t}$ at 8 epochs between 158 and 182d from the $ V$-band data, corrected for distance, extinction and a bolometric correction factor of $0.26 \pm 0.06$ mag during nebular phase \citep{2003ApJ...582..905H}. The weighted mean value of $L_{\rm t}$ is  $18.43\pm0.83\times10^{40}\,$\ergs, corresponding to a mean phase of 172d. This tail luminosity corresponds to a value of $M_{\rm Ni} =0.064\pm0.003$ \msun.

 \cite{2003A&A...404.1077E} has found a tight linear correlation between the $ Log (M_{\rm Ni})$ and the plateau $ V $-band absolute magnitude at ($ t_i-35 $) epoch, where $ t_i $ is inflection time.
 Which is defined as the moment when the slope of the light curve in the transition phase is maximum.
 For \sn\ light curve, we constrained $ t_i=102.99\pm0.02 $ d. Following the above mentioned correlation we obtain $M_{\rm Ni} =0.066\pm0.002$ \msun.

 We adopt the mass of synthesized \nickel\ in \sn\ to be $0.064\pm0.006 $ \msun, which is derived from the first method, and is found to be consistent with that obtained from the subsequent two methods described here. We anticipate that the estimated \nickel\ mass for \sn\ is almost equal to that obtained for SNe 2012aw, 2004et and 1999em. Whereas for sub-luminous SN 2005cs, \nickel\ mass is much less than \sn.

\section{Optical spectra} \label{sec:sp}

\subsection{Key spectral features} \label{sec:sp.key}

\begin{figure*}
\centering
\includegraphics[width=12cm]{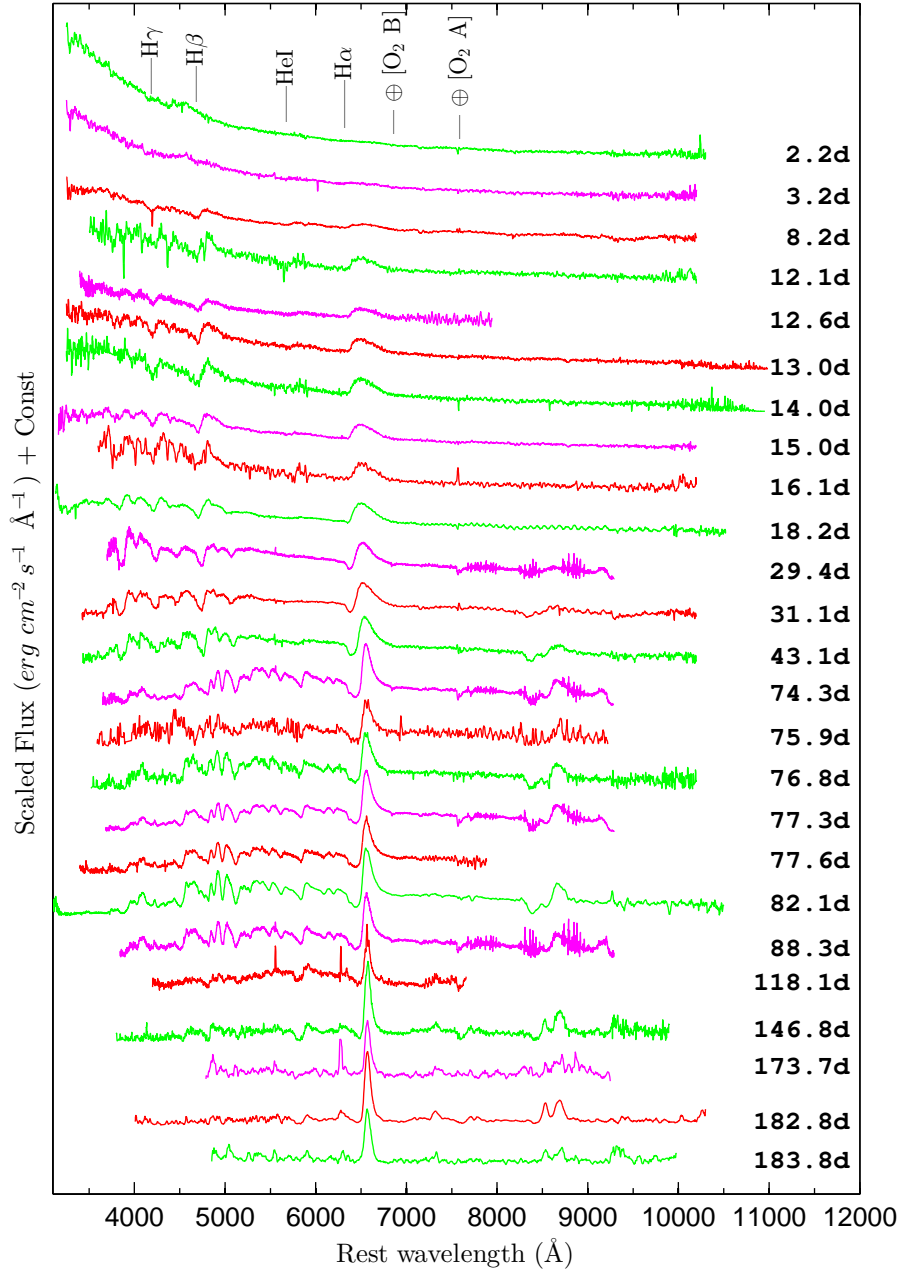}
\caption{The Doppler-corrected spectra of \sn\ are shown at 14 phases from 7d to 270d. Prominent P-Cygni
         profiles of hydrogen (\ha, \hb, \hg) and helium (\Hei\ $ \lambda $5876) lines are marked.  The telluric absorption features
         are indicated with the \earth\, symbol. The portions of spectra at the extreme blue and red ends have poor
         signal-to-noise ratios.}
\label{fig:sp.all}
\end{figure*}

The spectroscopic evolution of \sn\ is presented in Fig.~\ref{fig:sp.all}. Preliminary identification of spectral features has been done as in previous studies of IIP SNe \citep[e.g.][]{2002PASP..114...35L,2013MNRAS.433.1871B}. Early spectra, viz. 2.2d and 3.2d, shows featureless blue continuum with a broad and shallow P-Cygni dip detected near 4380 \AA\ which is supposedly \Heii\ $ \lambda $4686, which is blue shifted by about 19500 \kms. Detection of such \Heii\ features has been reported in several early SN spectra \citep{2001MNRAS.325..907F,2007ApJ...666.1093Q,2013A&A...555A.142I,2014arXiv1408.1404S,2015MNRAS.449.1921P}.
The 8.2d spectrum also primarily shows blue continuum, although with developing \hb, \Hei\ and \ha\ lines.
The \Hei\ feature completely disappears after 18d and at similar position \Nai~D profile starts to emerge since the 43d spectrum.

\begin{figure}
 \centering
 \includegraphics[width=8.4cm]{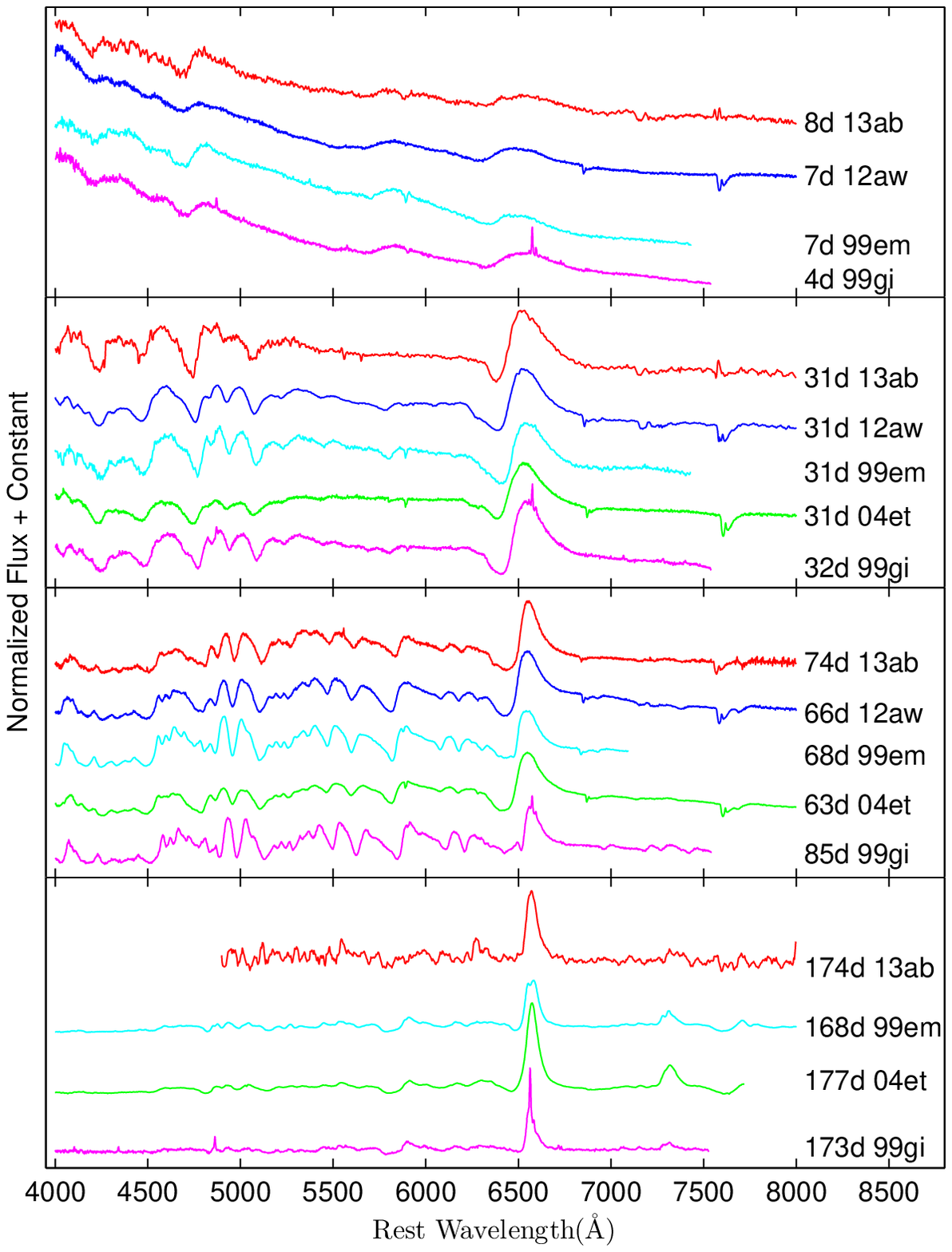}
 \caption{Comparison of early (8d),  plateau (31d, 74d) and nebular (174d) phase spectra of {\sn} with those of
          other well-studied type IIP SNe 2012aw \citep{2013MNRAS.433.1871B}, 1999em \citep{2002PASP..114...35L}, 1999gi \citep{2002AJ....124.2490L},
          2004et \citep{2006MNRAS.372.1315S,2010MNRAS.404..981M}. Observed fluxes of all
          the SNe are corrected for extinction and redshift (adopted values same as in Fig.~\ref{fig:lc.abs}).}
 \label{fig:sp.lit}
 \end{figure}

\begin{figure}
\centering
\includegraphics[width=8.5cm]{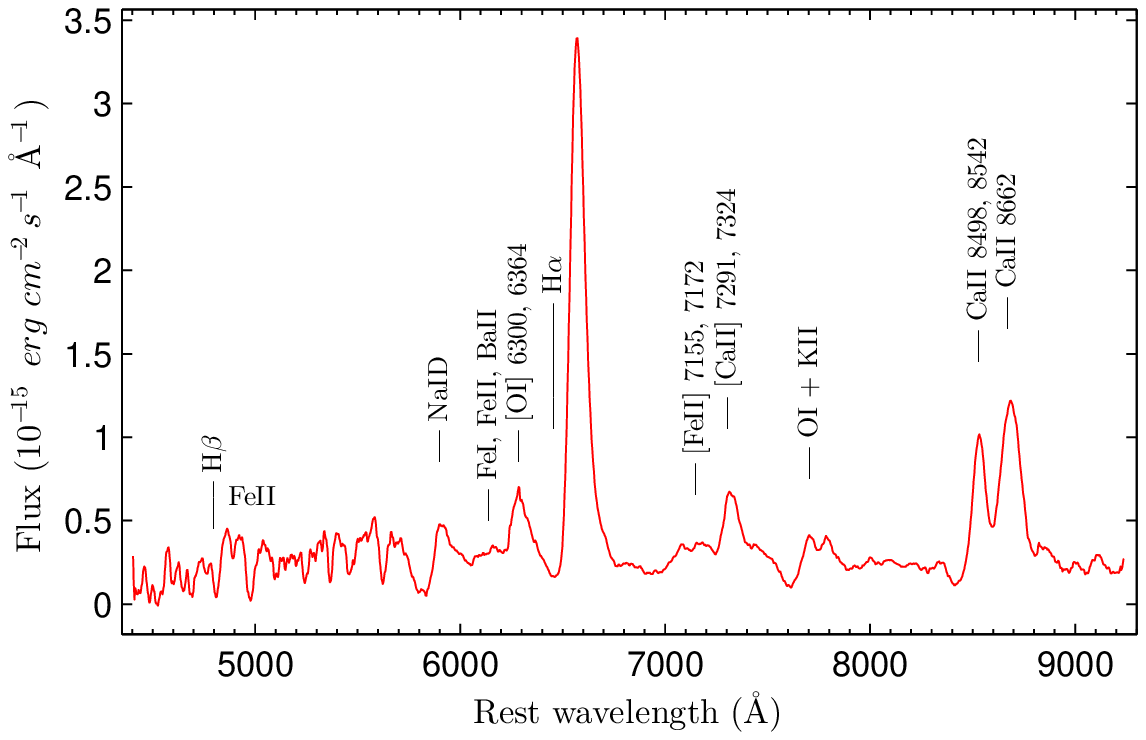}
\caption{Line identification in the nebular phase spectrum of \sn\ (183d).}
\label{fig:sp.neb}
\end{figure}

The spectra from 12 to 18d mark the transition phase from a hot to cool SN envelope, when photosphere begins to penetrate the deeper Fe-rich ejecta. These spectra mark the emergence of other lines from
heavier atomic species, such as  calcium, iron, scandium, barium, titanium and neutral sodium.
Among these lines,
\Feii~$ \lambda $5169 appear during the early plateau phase (12d), whereas weaker lines start to emerge at the beginning of late plateau phase (18d). \Nai~D doublet $ \lambda\lambda $~5890, 5896 and \Caii\ triplets $ \lambda\lambda $~8498, 8542, 8662 are feebly traceable from 31d, and becomes prominent since the 43d spectrum.
All weak and blended lines are seen to evolve and appear prominently by the end of plateau phase (82.1d). The following spectrum (88.3d) marks the onset of plateau to nebular transition.
All subsequent spectra up to 118.1d are representative of the early nebular phase, when the outer ejecta has become optically thin.
Fig.~\ref{fig:sp.lit} compares \sn\ spectra with sample of archetypal IIP events at four different epochs, viz. early and hot plateau phase at 8d; cooler plateau phase at 31 and 74d; and nebular phase at 174d. \sn\ spectra show features identical to those of our comparison sample of normal events. The nebular spectrum at 183d is shown in Fig.~\ref{fig:sp.neb} with preliminary identification of nebular lines typical  of SNe IIP. This spectrum is  mostly dominated by emission features of \Oia\ $ \lambda\lambda $~6300, 6364,
  \Caiia\ $ \lambda\lambda $~7291, 7324, and \Feiia\ $ \lambda\lambda $~7155, 7172. In addition, permitted emission lines of  \Hi, \Nai\ $ \lambda\lambda $~5890, 5896 doublet and the \Caii\ NIR triplet are still detected.

\subsection{\textsc{SYNOW} modelling of spectra} \label{sec:synow}

Spectra of \sn\ have been modeled with \texttt{\synow\ 2.3}\footnote{http://www.nhn.ou.edu/$\sim$parrent/synow.html} \citep{1997ApJ...481L..89F,1999MNRAS.304...67F,2002ApJ...566.1005B} for preliminary line identification and velocity estimates.
\synow\ is a parametrized spectrum synthesis code which employs Sobolev approximation to simplify radiation transfer equations, assuming spherically symmetric supernova ejecta which expand homologously. Despite of the simplified LTE atmosphere assumption in \synow, it is capable to produce P-Cygni profiles like those produced in the expanding photosphere of a SN. \synow\ has been implemented on several recent SNe studies \citep[e.g.][]{2012MNRAS.422.1178I,2013MNRAS.433.1871B,2013ApJ...767...71M,2014ApJ...782...98B, 2014MNRAS.438..368T,2014ApJ...781...69M} for line identification and estimation of line velocities.
We tried three different options for optical depth profiles (viz. Gaussian, exponential
and power law), no significant differences were noticed. However, while matching absorption minimum, the exponential profile,
$\tau\propto exp(-v/v_e)$, where $v_{e}$, a profile fitting parameter, e-folding
velocity, was found to be the most suitable and is adopted here for each individual atomic species.
One important aspect of \synow\ modeling is the concept of detachment of an ion. When the minimum velocity of a line-forming layer is higher than that of the photospheric layer, the ion is said to be detached, which results into flat topped emission and blueshifted absorption counterpart of the line profile in synthetic spectra produced by \synow. This becomes important for \ion{H}{I} lines as they are essentially formed at much higher velocities than photospheric velocities. Therefore, only the detached scenario for \ion{H}{I} reliably fits the blue-shifted absorption trough in the observed spectra.

\begin{figure*}
\centering
\includegraphics[width=\linewidth]{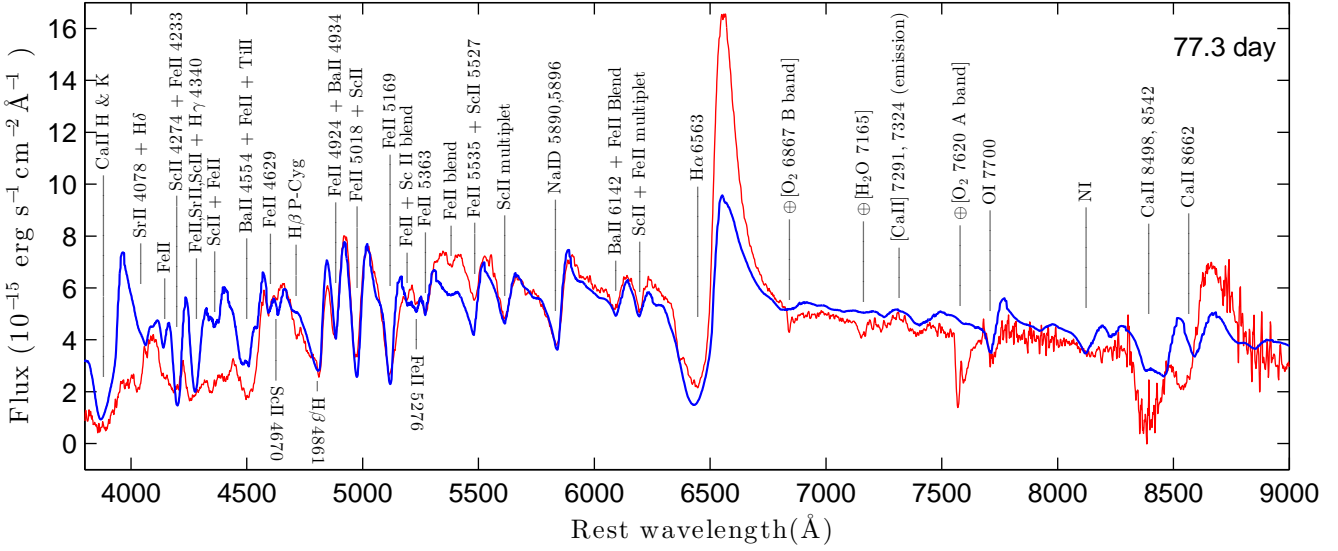}
\caption{\synow\ modelling of the 77.3d spectrum of \sn. A \synow\ model spectrum is shown with thick solid line (blue),
         while the observed one is the thin solid line (red). Fluxes are corrected for interstellar extinction.}
\label{fig:sp.synph}
\end{figure*}

The observed spectra are dereddened  and doppler-corrected before modelling with \synow. $T_{bb}$ is supplied as a model input parameter which is actually the blackbody temperature to produce the underlying LTE continuum of synthetic spectra. For this reason, the observed spectral continuum is well matched at early phases, whereas at later phases this is a poor match with the model as SN emission significantly deviates from LTE assumption. Fig.~\ref{fig:sp.synph} shows the observed spectrum at 77.3d with our best-fit model.
A set of atomic species (\Hi, \Hei; \Feii; \Tiii; \Scii; \Caii; \Baii; \Nai; \Siii; \Oi; \Ni) has been incorporated to generate the synthetic spectrum.
The model spectrum can very well reproduce most of the blended line profiles; $v_{ph}$ is optimized to match the \Feii\ multiplet ($ \lambda\lambda $~4924, 5018, 5169), whilst the \Hi\ velocities are always dealt as a detached scenario.

\begin{table*}
  \caption{The best-fit blackbody continuum temperature ($T_{bb}$) and line velocities
           of \hb, \Feii\ ($ \lambda\lambda $~4924, 5018, 5169) and \Hei\ $ \lambda $5876 as estimated
           from \synow\ modeling of the observed spectra of \sn. Velocities derived
           using lines of \Feii\ or \Hei\ are taken as representative of the velocity of photosphere ($v_{\rm ph}$).}
  \label{tab:synow}

  \begin{tabular}{ccccccc } \hline
       UT Date     &Phase$^{a}$ & $T_{bb}^{b}$ & $v$(\Hei)      & $v$(\Feii)  & $v$(H$_\beta$)\\
     (yyyy-mm-dd)  &   (day)    & (kK)         &$10^{3}$ \kms   & $10^{3}$\kms& $10^{3}$\kms  \\ \hline
    2013-02-24.71  &   8.21     & 11.8         & 10.2$\pm$0.3   &   -         & 10.4$\pm$0.6  \\
    2013-02-28.59  &  12.09     & 9.6          & 10.1$\pm$0.7   &   -         & 10.4$\pm$0.6  \\
    2013-03-01.08  &  12.58     & 1.1          & 9.4 $\pm$0.4   &   -         & 10.3$\pm$0.3  \\
    2013-03-01.50  &  13.00     & 9.9          & 9.0 $\pm$0.5   &   -         & 10.3$\pm$0.2  \\
    2013-03-02.51  &  14.01     & 1.0          & 8.7 $\pm$0.9   &   -         & 10.2$\pm$0.5  \\
    2013-03-03.51  &  15.01     & 9.2          & 9.0 $\pm$0.6   &   -         &  9.5$\pm$0.2  \\
    2013-03-04.59  &  16.09     & 8.4          & 8.2 $\pm$0.8   &   -         &  9.3$\pm$0.8  \\
    2013-03-06.71  &  18.21     & 8.5          & 8.3 $\pm$0.6   &   -         &  9.8$\pm$0.4  \\
    2013-03-17.93  &  29.43     & 6.8          &  -             &  6.1$\pm$0.3&  7.0$\pm$0.3  \\
    2013-03-19.63  &  31.13     & 5.9          &  -             &  6.5$\pm$0.4&  7.5$\pm$0.2  \\
    2013-03-31.58  &  43.08     & 5.6          &  -             &  4.9$\pm$0.4&  6.0$\pm$0.3  \\
    2013-05-01.84  &  74.34     & 5.0          &  -             &  3.1$\pm$0.3&  3.1$\pm$0.4  \\
	2013-05-03.35  &  75.85     & 5.2          &  -             &   -         &  -            \\
	2013-05-04.33  &  76.83     & 5.1          &  -             &  3.4$\pm$0.4&  3.8$\pm$0.3  \\
	2013-05-04.85  &  77.35     & 5.2          &  -             &  2.9$\pm$0.3&  3.0$\pm$0.3  \\ \hline

  \end{tabular}
\begin{flushleft}
  $^{a}$ With reference to the time of explosion JD 2456340.0\\
  $^{b}$ Best-fit blackbody temperature at the photosphere to match the continuum in the observed spectrum.\\
\end{flushleft}
\end{table*}

Line velocities for \hb, \Hei\ and \Feii\ are estimated from all spectra.
The model fit is optimized for velocity locally around the respective lines of interest. Fitting is done locally to avoid any bias in velocity estimation, which may be imposed while accounting for entire spectrum due to other lines at different velocities.
In Table~\ref{tab:synow}, \synow\ estimated velocities for a few representative lines in the plotted spectral sequence are shown.
Although \Feii\ line impression is detectable since 12d, we model those lines only from 29d onwards, as in earlier spectra \Feii\ triplet is either not full developed or only detectable \Feii\ $ \lambda $5169 is too weak to model.
\textsc{synow} modelling is done until 77d, because after 78d spectra are limited because of low signal-to-noise ratio (SNR).
In such low SNR spectra \synow\ may not provide any better estimation of line velocities than absorption-minima position measurements.

\subsection{Evolution of spectral lines} \label{sec:sp.line}

\begin{figure}
 \centering
 \includegraphics[height=15.40cm]{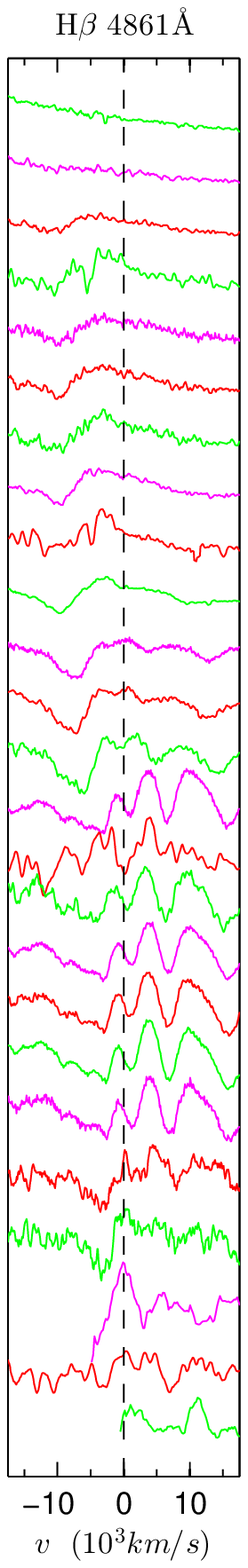} \hskip -4mm
 \includegraphics[height=15.51cm]{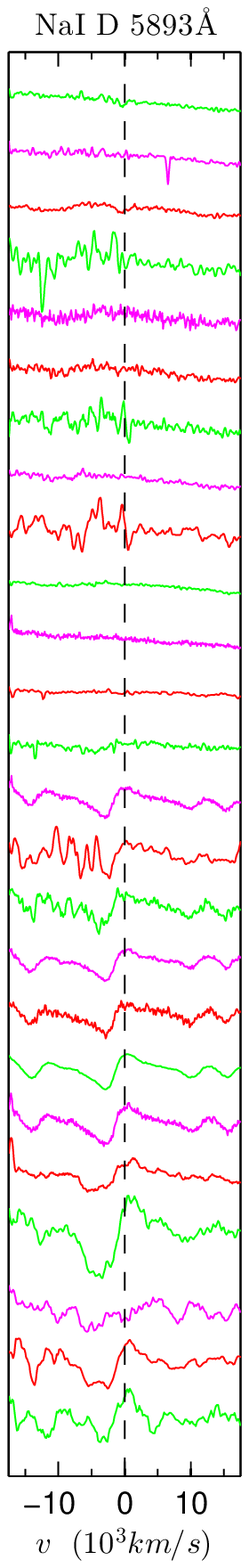} \hskip -4mm
 \includegraphics[height=15.51cm]{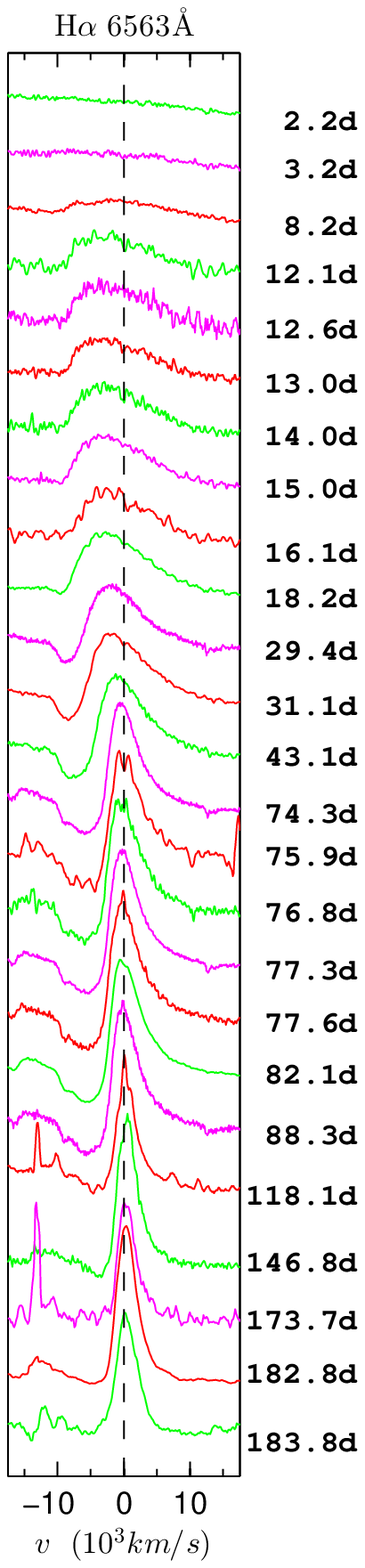}
 \caption{Evolution of the line profiles of \hb, \Nai~D and \ha\ plotted at 25 epochs from 2d to 184d.
 The zero-velocity position is shown with a dotted line and the corresponding rest
          wavelength is indicated at the top.}
 \label{fig:sp.line}
 \end{figure}

 The evolution of spectral features provides important clues about the interaction of expanding ejecta
 with the circumstellar material, formation of dust in the ejecta and geometrical distribution of
 the ejecta. To illustrate the evolution of individual lines, in Fig.~\ref{fig:sp.line} selected  regions of spectra are plotted in the velocity domain corresponding to the rest wavelengths of \hb, \Nai~D and \ha. There is no clear evidence of spectral lines in the early spectra (2.2 and 3.2d), except for a shallow and broad \Heii\ ($ \lambda $4686) feature near 4380\AA.
 The blue-shifted absorption troughs of the P-Cygni profiles give direct estimate of expansion velocity of the ejecta. The emission peaks are also seen to be blue-shifted. The amount of blue-shift decreases with the decline of the expansion velocity and settles to the rest velocity while the SN enters the nebular phase.
 This is a generic feature seen in SN spectra, mostly at early phases; see, e.g., SNe 1987A \citep{1987A&A...182L..29H}, 1998A \citep{2005MNRAS.360..950P}, 1999em \citep{2003MNRAS.338..939E}, 2004et \citep{2006MNRAS.372.1315S} and 2012aw \citep{2013MNRAS.433.1871B}. Blue-shifted emission peaks are explained by the diffused reflection of photons from expanding SN envelope \citep{1988SvAL...14..334C} which is in contrast to the pure-absorption model of expanding atmosphere ensuing un-shifted emission peaks. However, recent study by \cite{2014MNRAS.441..671A} suggests that these features are tied with the density structure of ejecta which in turn controls the amount of occultation of the receding part of ejecta, resulting in biasing of the emission peak. Such features are well reproduced by  non-LTE models like \cmfgen\ \citep{2005ASPC..332..415D}. The evolution of blue-shifted peaks are clearly seen in \ha, whilst other lines emission peaks are heavily contaminated by P-Cygni the absorptions from other adjacent lines.
 The emission peak blue-shift for \ha\ is found to be as high as $ \sim-5000 $\kms at 8.2d and then progressively decreases (4400 \kms\ at 12d and 2300 \kms at 29.4d) down to almost zero velocity at 88.3d, corresponding to the end of plateau phase.

Similar to \ha, \hb\ are seen to evolve all throughout the spectral evolution (see Fig.~\ref{fig:sp.line}). However, 18d onwards, the red side of \hb\ emission profiles are found to be significantly dominated by emerging \Feii\ lines. All three \Feii\ lines ($ \lambda\lambda $~4924, 5018, 5169) are seen to have fully appeared in 31d spectrum, which continues to evolve till the last observation. Traces of \Hei\ line are seen in early spectra, which disappears after 18d. At similar position \Nai~D profile start to appear at 43d and it continues to evolve until last spectrum.

\subsection{Ejecta velocity} \label{sec:sp.vel}

The element distribution of the progenitor at the end of the sequence of nuclear burnings and before the SN explosion is stratified, with
 hydrogen being abundant in outermost layers and heavier elements (e.g. $ \alpha $-elements) towards the center and core being rich in iron. Thus it is expected that the expanding ejecta are constituted by layers of multiple elements and the so called ``onion-like" structure. Therefore, different lines originate from different depths in the SN atmosphere.
 The photosphere is a region of special interest to study the kinematics and other related properties. The photosphere represents the layer of SN atmosphere where optical depth attains a value of $\sim~^2/_3 $ \citep{2005A&A...437..667D}. No single spectral line can represent the true photospheric layer and its velocity.
 During the plateau phase, \Feii\ ($ \lambda\lambda $~4924, 5018 and 5169) or \Scii\ lines are thought to be the best estimators for the photospheric velocity ($v_{\rm ph}$), and at early phases when \Feii\ lines are not strongly detected, the best proxy for $v_{\rm ph}$ is \Hei\ or even \hb\ \citep{2012MNRAS.419.2783T}, at earlier phases.

 Velocities can be estimated either by simply locating the blue-shifted absorption trough of the P-Cygni profiles or modeling the observed spectra where velocity is one of the input parameter. We have used both methods to estimate the velocities.
SN spectral lines are often found to be blended with other lines in the neighborhood, as in case of \Feii\ multiplets, blended with \Tiii\ and \Baii. This introduces some error, when lines velocities are estimated by locating the absorption minima, by simply fitting gaussian function on these blended profiles, which is further exacerbated in low-resolution and low-SNR spectra. \synow\ being capable to reproduce P-Cygni profile for multiple lines of different ions simultaneously, it can easily reproduce line blending as observed in SN spectra. Thus we get a better handle while fitting the entire blended profile with \synow\, and so the velocity estimates are better and less prone to errors. More detailed discussion on applicability  and merits of \synow\ velocity estimates over absorption minima method can be found in \cite{2012MNRAS.419.2783T} and \cite{2014ApJ...782...98B}.

In order to estimate photospheric velocities from the model, \synow-generated synthetic spectra are locally fitted over \Feii\ lines in the observed spectra. This was done to avoid any over- or under- estimation of velocities as each line of any atomic species originating from different layers.
 The attributed uncertainties are visually estimated by noting the deviation of model absorption trough from observed one.
 The model velocities listed in Table~\ref{tab:synow} are estimated only until 77.3d, since thereafter the spectra are limited by SNR and hence there is no advantage in using \synow-estimated velocities over the absorption minima method.
 The \Feii\ velocities are estimated by assuming that the \Feii\ $ \lambda $4924, \Feii\ $ \lambda $5018, and \Feii\ $ \lambda $5169 lines have velocities coincident with $v_{\rm ph}$, whereas the Balmer \Hi\ lines are treated as detached (with $ \rm v > v_{\rm ph}$) \citep{2002ApJ...566.1005B,2013MNRAS.433.1871B}.
 \Hei\ line velocities have also been estimated as long as this ion can be traced in spectra (from 8.2 to 18.2 d).

\begin{figure}
 \includegraphics[width=8.5cm]{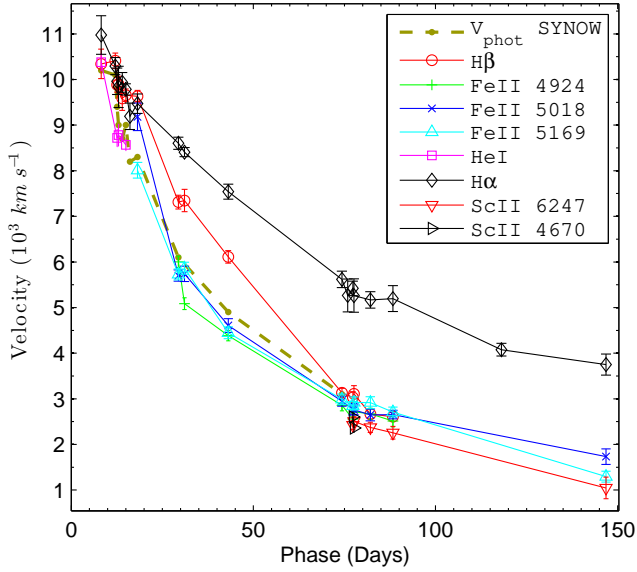}
 \caption{Evolution of the \ha, \hb, \Hei, \Scii\ and \Feii~ line velocities. The velocities are estimated through the
          Doppler-shift of the absorption minima. The expansion velocities at the photosphere ($v_{\rm ph}$) estimated from
          \synow\ fits of \Hei\ line until 18d and simultaneous fits for \Feii\ lines at later phases
          (see Table~\ref{tab:synow}) are shown as a comparison.}
 \label{fig:sp.velall}
 \end{figure}

 The expansion velocities of \ha, \hb, \Hei, \Feii\ ($ \lambda\lambda $~4924, 5018 and 5169),
 \Scii\ $ \lambda $6247 and \Scii\ $ \lambda $4670
 have also been determined using \iraf\, by fitting the absorption trough with a Gaussian
 profile. The results are plotted in Fig.~\ref{fig:sp.velall}. It can be seen that \Hi\ lines are formed at larger radii than \Hei\, whereas \Feii\ lines are formed at much smaller radii, having lower velocities. The \Scii\ lines are formed at a even smaller radius.
 \Scii\ lines are weak in strength and due to the limitation of our low SNR spectra, these lines are only detected during 77 to 147d.

\begin{figure}
\centering
\includegraphics[width=8.5cm]{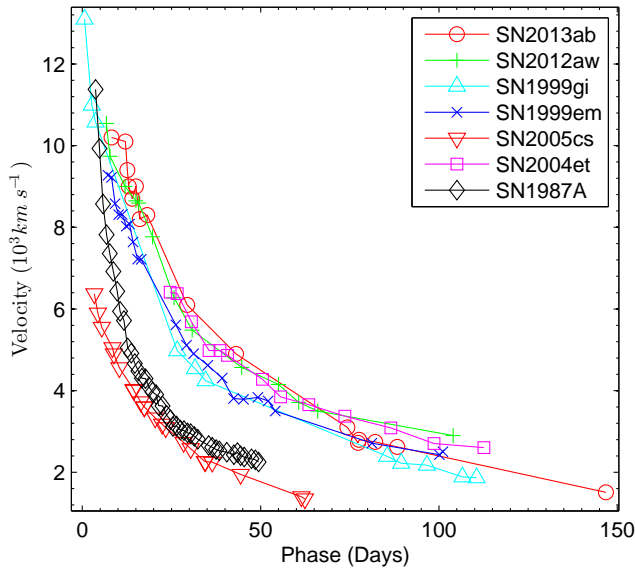}
\caption{The evolution of photospheric velocity ($v_{\rm ph}$) of {\sn} is compared with those of other well-studied SNe.
         The $v_{\rm ph}$ plotted here are the absorption trough velocities (from \Hei\ lines at early phases,  and \Feii\ at late phases.)}
\label{fig:sp.velph}
\end{figure}

 Fig.~\ref{fig:sp.velph} shows the comparison of photospheric velocity of \sn\ with those of other well-studied
 SNe 1987A, 1999em, 1999gi, 2004et, 2005cs and 2012aw.
 For the purpose of comparison, the absorption trough velocities have been used, taking the mean of \Feii\ lines (or \Hei\ lines at early phases where \Feii\ lines are not detected). The velocity profile of \sn\
 is very similar to those of other normal SNe IIP. On the other hand, the velocities of SN 2005cs and SN 1987A are significantly smaller than those of normal events including SN 2013ab. The shape and values of the \sn\ velocity profile is strikingly similar to those of SNe 2004et and 2012aw, whereas the velocities are consistently higher than SNe 1999gi and 1999em by $ \sim800 $\kms.

 \begin{figure}
 \centering
 \includegraphics[width=8.4cm]{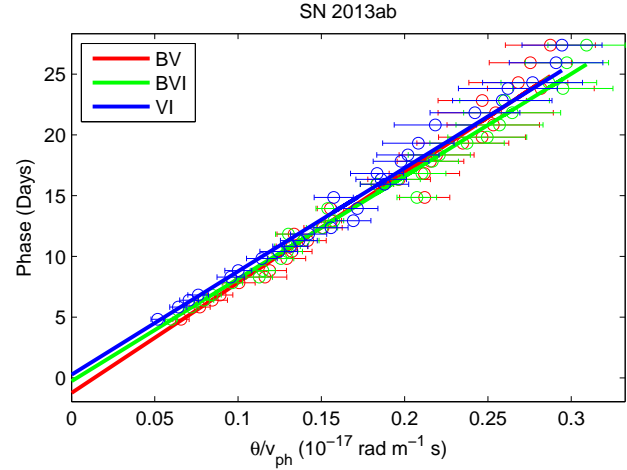}
 \caption{The EPM fit for \sn,  using \citet{2005A&A...437..667D} prescription for the dilution factor and using three filter sets.}
 \label{fig:epm}
 \end{figure}

\section{Expanding Photosphere Method} \label{sec:epm}
Expanding photosphere method (\epm) is a geometrical technique \citep{1974ApJ...193...27K,1996ApJ...466..911E} in which the angular radius is compared with physical radius of the SN photosphere to estimate its distance. Assuming homologous expansion of SN photosphere, the physical radius at any instant is approximated from expansion velocity($ v_{ph} $), and angular radius ($ \theta $) is estimated from blackbody fit corrected with the dilution factor ($ \xi $) of \cite{2005A&A...437..667D} for non-LTE SNe atmosphere.
Distance ($ D $) and explosion epoch ($ t_0 $) are related with the quantity $ \theta/v_{ph} $ at any given time $ t $ as,
\begin{equation}
t=D\left(\frac{\theta}{v_{ph}}\right)+t_0
\end{equation}
\noindent Thus, the plot of $ t $ against $ \theta/v_{ph} $, yield distance as the slope and explosion epoch as the y-intercept.

EPM has been successfully applied to a considerably large sample of type IIP events by \cite{2009ApJ...696.1176J} and \cite{2014ApJ...782...98B}. The merits and limitations of the method has also been tested for multiple aspects. Here, we followed the same approach as described in \cite{2014ApJ...782...98B}.
It has also been shown that the two dilution factor prescriptions given by \cite{2001ApJ...558..615H} and \cite{2005A&A...439..671D} have significant differences with the latter being more consistent with other redshift-independent distance estimates. In addition also the \synow-estimated velocities are better suited for such analysis.
Taking into account all these factors,  we have used the \cite{2005A&A...439..671D} dilution factor prescription and \synow\ estimated velocities for \epm\ analysis. We also restricted our data set up to 30d only.
We derived an \epm\ distance of $ 24.26\pm0.98 $ Mpc which is the mean value for $ BV $, $ BVI $ and $ VI $ band-sets (Fig.~\ref{fig:epm}). The corresponding derived explosion epoch from \epm\ is JD $ 2456339.59\pm0.76 $ day,
which is in good agreement with the uncertainty of adopted explosion epoch from observations (see \S\ref{sec:intro}). In principle we can constrain explosion epoch and keep distance as the only free parameter for the analysis. This yields a distance of $ 23.49\pm0.77 $ Mpc. Since, the derived explosion epoch from unconstrained analysis is within the uncertainty of the observationally adopted explosion epoch, and the derived distances from constrained and unconstrained analysis are reasonably consistent within the limits of uncertainty, we prefer to adopt the \epm\ distance as derived from the unconstrained analysis.
The \epm\ distance is found to be in good agreement with mean redshift independent
distance estimates ($ 24.9\pm5.8 $ Mpc) listed in NED\footnote{NASA Extragalactic Database http://ned.ipac.caltech.edu/} database for the galaxy \host.

\section{Characteristics of the explosion} \label{sec:char}

\subsection{Radius of progenitor} \label{sec:radius}

\begin{figure}
\centering
\includegraphics[width=8.5cm]{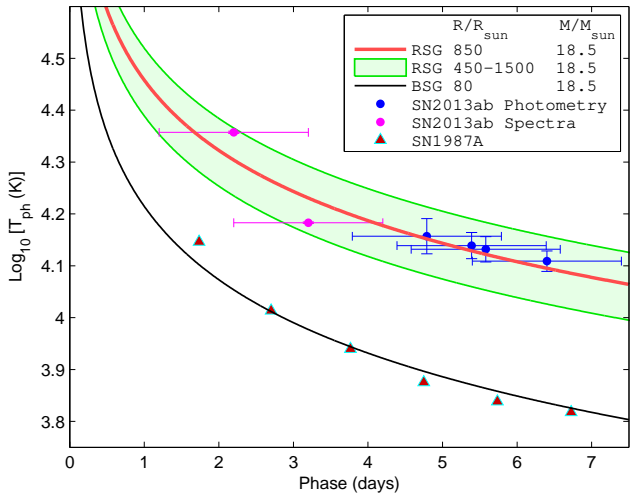}
\caption{Constraining radius using \citet{2011ApJ...728...63R} prescription. The red solid line is the best fit for RGS of 850 \rsun\ for \sn\ temperatures and the black solid line is for BSG of 80 \rsun\ for SN 1987A temperatures \citep{1987MNRAS.227P..39M}.}
\label{fig:rad.const}
\end{figure}

 \begin{figure*}
 \centering
 \includegraphics[height=0.75\linewidth, angle=-90]{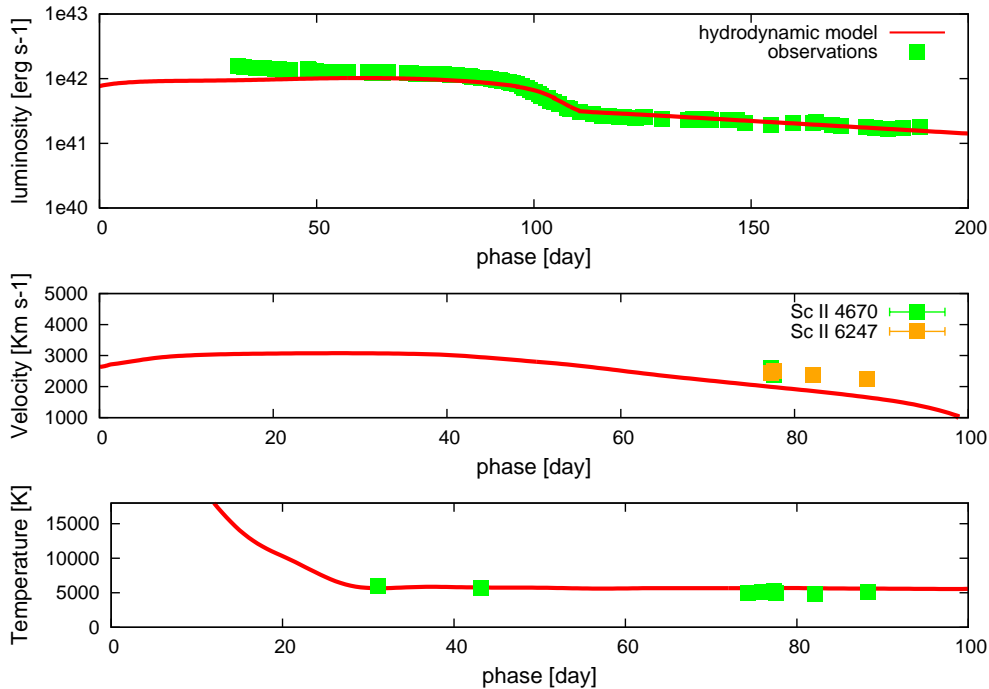}
 \caption{Comparison of the evolution of the main observables of SN 2013ab with the best-fit model computed using the general-relativistic,
 radiation-hydrodynamics code (total energy $0.35$ foe, initial radius $4.2 \times 10^{13}$ cm,
 envelope mass $7\,{\rm M_{\sun}}$). Top, middle, and bottom panels show the bolometric light
 curve, the photospheric velocity, and the photospheric temperature as a function of time. To better
 estimate the photosphere velocity from observations, we use the minima of the profile of the
 \Scii\ lines
 which
 are considered good tracer of the photosphere velocity in Type II SNe. As for the photospheric temperature,
 we use the blackbody temperature derived from the blackbody fits to the spectral continuum.
 }
 \label{fig:model}
 \end{figure*}

During the shock breakout CCSNe heats the envelope to extremely high temperatures and then it cools down as the envelope expands. This temperature evolution depends on several parameters, viz. progenitor radius, opacity, explosion energy and mass. A simplified analytic formulation relating all these parameters has been determined by \cite{2011ApJ...728...63R}.  For a progenitor with larger radius, the envelope remains at a higher temperature for a longer time as compared to a progenitor with smaller radius.
The temperature evolution is weakly dependent on the progenitor mass and energy, whereas radius and opacity are the dominating parameters. We computed blackbody temperatures from photometric fluxes and spectra and constrained progenitor radius by fitting the relation on these values (Fig.~\ref{fig:rad.const}). We restricted our fits only within one week of the explosion as the relation is valid for the initial few days after the explosion. We adopted an optical opacity of 0.34 and a typical RSG density profile $ f_p=0.11 $, however it is also noted that results are not very sensitive towards $ f_p$. This analysis yields an initial radius of 750-950 \rsun\ for the progenitor of \sn,  adopting minimal line-of-sight extinction (see \S\ref{sec:ext}).
The radius estimate suggests that the progenitor is possibly a large RSG. With increased extinction value the estimated radius would increase even further.

\subsection{Hydrodynamical modelling}\label{modelling}

With the same well-tested approach adopted for other observed SNe \citep[e.g.
SNe 2007od, 2009E, 2009N, 2012A, and 2012aw; see][]{2011MNRAS.417..261I,2012A&A...537A.141P,2014MNRAS.438..368T,2013MNRAS.434.1636T,2014ApJ...787..139D},
we constrain the
main physical properties of SN 2013ab at the explosion (namely the ejected mass, the
progenitor radius and the explosion energy) through the hydrodynamical modelling of
the main observables (i.e. bolometric light curve, evolution of line velocities and continuum
temperature at the photosphere).

According to this approach, a simultaneous $\chi^{2}$ fit of the above mentioned observables
against model calculations is performed. Two codes are employed for the computation of the
models: 1) the first one is the semi-analytic code where the energy balance equation is solved
for a homologously expanding envelope of constant density \citep[for details see][]{2003MNRAS.338..711Z};
2) the other one is the new general-relativistic, radiation-hydrodynamics Lagrangian code presented
in \cite{2010MSAIS..14..123P} and \cite{2011ApJ...741...41P}, which is able to simulate the
evolution of the physical properties of the CC-SN ejecta and the behavior of the main observables
from the breakout of the shock wave at the stellar surface up to the nebular stage. The distinctive
features of this new code are \citep[cf. also][]{2013MNRAS.434.3445P}: a) an accurate treatment of radiative
transfer coupled to hydrodynamics, b) a fully implicit Lagrangian approach to solve the coupled non-linear finite difference system of general-relativistic, radiation-hydrodynamics
equations, and c) a description of the evolution of ejected material which takes into account both
the gravitational effects of the compact remnant and the heating effects linked to the decay of
the radioactive isotopes synthesized during the SN explosion.

The semi-analytic code  used to carry out a preparatory study aimed at constraining the
parameter space describing the SN progenitor at the explosion. The results of such study
are exploited to guide the more realistic, but time consuming model calculations performed with
the general-relativistic, radiation-hydrodynamics code.

We note that modelling with both codes is appropriate, since the emission of SN 2013ab is dominated
by the expanding ejecta.
However, in performing the
$\chi^{2}$ fit, we do not include the observational data taken at early phases (first $\sim$ 20-30 days
after explosion) because our model is not able to accurately reproduce the early evolution
of the main observables, due to the approximate initial density profile used in the simulations which
does not precisely reproduce the radial profile of the outermost high-velocity shell of the ejecta
forming after the breakout of the shock wave at the stellar surface \citep[cf.][]{2011ApJ...741...41P}.

The explosion epoch (JD $= 2456340.0$) and distance modulus
($\mu= 31.90$ mag) adopted in this paper (see \S~\ref{sec:ext}) are used to fix the explosion epoch
and to compute the bolometric luminosity of SN 2013ab, both necessary to perform the comparison
with model calculations.
Since we do not have infrared (IR) observations for \sn, we computed true bolometric luminosity by adding IR contribution, assuming \sn\ has similar optical to IR flux ratio as observed for SN 1999em.
Adopting a \nickel\ mass of $0.06\,{\rm M_{\sun}}$ (see \S~\ref{sec:lc.nick}), the best fit procedure
returns values of kinetic plus thermal energy of 0.35 foe, initial radius of $ 4.2\times10^{13} $ cm ($ \sim600 $~\rsun)
and envelope mass of 7~\msun\ (see Fig.~\ref{fig:model}
), with an estimated
uncertainty on the modelling parameters of about 15\%. Adding the mass of the compact remnant ($\sim 1.5-2.0\,{\rm M_{\sun}}$) to that of the ejected material, the mass of
the progenitor of SN 2013ab at the explosion would be $\sim 9~\msun$.

From the middle panel of Fig.~\ref{fig:model} one can note a small ($ ~ $15-20\%) discrepancy between the observed photospheric velocity and our best-fitting model. Such discrepancy may be linked to a systematic shift between the true photospheric velocity and the values estimated from the observed P-Cygni line profiles \citep{2005A&A...439..671D},
according to which the optical depth in the lines could be higher than that in the continuum, causing a shift
of the line photosphere to a larger radius \citep[see also][]{2013A&A...555A.142I}. Nevertheless we notice that the discrepancy can be eliminated by adopting for the model higher values for the initial radius ($ \sim 6\times10^{13} $ cm),
the  energy ($ \sim 0.6 $ foe) and the envelope mass (up to 13 \msun). However in this case we get a worse
fit to the observed light curve, with a longer (by about 30\%) plateau.

The values reported above are consistent with a core collapse scenario from a typical red super-giant
progenitor of relatively low mass. The radius estimate is also consistent within errors with that estimated from early photospheric temperatures described in \S\ref{sec:radius}.

\section {Summary} \label{sec:sum}
In this paper we present high-cadence broadband photometric and low resolution spectroscopic observations of \sn\ spanning a duration of about 6 months. In total, we collected 135 epoch of photometric and 25 epochs of spectroscopic observations.
A brief summary of the results obtained in this work is given below.

\begin{enumerate}
\item The light curve and the bolometric luminosity comparisons with other SNe IIP suggest that \sn\ is a normal SN IIP, though with a relatively large plateau decline rate (0.92 mag 100 d$ ^{-1} $ in the $ V$ band) and a shorter plateau duration ($ \sim 78 $ d). The $^{56}$Ni mass estimated by comparing the tail luminosity with that of SN 1987A yields a value of 0.064~\msun.

\item
Spectroscopic comparisons show strong resemblance with canonical type IIP events. Earliest spectra show a featureless continuum. As the SN evolve, the spectra develop metal lines (calcium, iron, scandium, barium, titanium and neutral sodium). Nebular phase spectra show emission lines with little or no P-Cygni signatures. Spectra until 77d with good signal-to-noise ratio are modeled using \synow\ to identify most of the prominent features and to estimate expansion velocities for the  \Hei, \Feii\ and \hb\ lines.

\item The \textsc{EPM} has been applied to \sn\ using the \synow-derived velocities and $ BVI $ photometric data. This provides an independent and reliable estimate of the distance of the galaxy \host\ as $ 24.3\pm1.0 $ Mpc.

\item We constrained the physical properties of \sn\ at the explosion by means of a hydrodynamical modelling of the main observables which uses the general-relativistic, radiation-hydrodynamics code described in \cite{2011ApJ...741...41P}. The kinetic plus thermal energy is estimated to be $ \sim0.35 $~foe, the progenitor mass is $ \sim9 $~\msun\ and the radius is about 600 $~\rsun$.

\end{enumerate}

\section*{Acknowledgments}
This work makes use of observations from the LCOGT network.
We are thankful to the observing and technical staffs of LCOGT, ARIES, HCT and Asiago telescopes for their kind cooperation in observation of \sn. We gratefully acknowledge the services of the
NASA ADS and NED databases and also the online supernova spectrum archive (SUSPECT) which are used to access data and references in this paper.
We acknowledge the TriGrid VL project and the INAF-Astronomical Observatory of Padua for
the use of computer facilities. M.L.P. and A.P.~acknowledges the financial support from CSFNSM and from the PRIN-INAF 2011 ``Transient Universe: from ESO Large to PESSTO'' (P.I.~S.~Benetti). We are also thankful to the anonymous referee, whose comments and suggestions has helped in significant improvement of the manuscript.

\bibliography{ms}

\begin{thebibliography}{82}
\providecommand{\natexlab}[1]{#1}

\bibitem[{{Anderson} et~al.(2014{\natexlab{a}})}]{2014MNRAS.441..671A}
{Anderson} J.~P. et~al., 2014{\natexlab{a}}, \mnras, 441, 671

\bibitem[{{Anderson} et~al.(2014{\natexlab{b}})}]{2014ApJ...786...67A}
{Anderson} J.~P. et~al., 2014{\natexlab{b}}, \apj, 786, 67

\bibitem[{{Arnett}(1996)}]{1996snih.book.....A}
{Arnett} D., 1996, {Supernovae and Nucleosynthesis: An Investigation of the
  History of Matter from the Big Bang to the Present}

\bibitem[{{Arnett}(1980)}]{1980ApJ...237..541A}
{Arnett} W.~D., 1980, \apj, 237, 541

\bibitem[{{Barbon} et~al.(1990){Barbon}, {Benetti}, {Rosino}, {Cappellaro} \&
  {Turatto}}]{1990A&A...237...79B}
{Barbon} R., {Benetti} S., {Rosino} L., {Cappellaro} E., {Turatto} M., 1990,
  \aap, 237, 79

\bibitem[{{Bayless} et~al.(2013)}]{2013ApJ...764L..13B}
{Bayless} A.~J. et~al., 2013, \apjl, 764, L13

\bibitem[{{Bersten} et~al.(2011){Bersten}, {Benvenuto} \&
  {Hamuy}}]{2011ApJ...729...61B}
{Bersten} M.~C., {Benvenuto} O., {Hamuy} M., 2011, \apj, 729, 61

\bibitem[{{Blanchard} et~al.(2013)}]{2013CBET.3422....1B}
{Blanchard} P. et~al., 2013, Central Bureau Electronic Telegrams, 3422, 1

\bibitem[{{Bose} \& {Kumar}(2014)}]{2014ApJ...782...98B}
{Bose} S., {Kumar} B., 2014, \apj, 782, 98

\bibitem[{{Bose} et~al.(2013)}]{2013MNRAS.433.1871B}
{Bose} S. et~al., 2013, \mnras, 433, 1871

\bibitem[{{Branch} et~al.(2002)}]{2002ApJ...566.1005B}
{Branch} D. et~al., 2002, \apj, 566, 1005

\bibitem[{{Breeveld} et~al.(2011){Breeveld}, {Landsman}, {Holland}, {Roming},
  {Kuin} \& {Page}}]{2011AIPC.1358..373B}
{Breeveld} A.~A., {Landsman} W., {Holland} S.~T., {Roming} P., {Kuin} N.~P.~M.,
  {Page} M.~J., 2011, in J.E. {McEnery}, J.L. {Racusin}, N.~{Gehrels}, eds,
  American Institute of Physics Conference Series. American Institute of
  Physics Conference Series, Vol. 1358, pp. 373--376

\bibitem[{{Brown} et~al.(2009)}]{2009AJ....137.4517B}
{Brown} P.~J. et~al., 2009, \aj, 137, 4517

\bibitem[{{Brown} et~al.(2013)}]{2013PASP..125.1031B}
{Brown} T.~M. et~al., 2013, \pasp, 125, 1031

\bibitem[{{Chugai}(1988)}]{1988SvAL...14..334C}
{Chugai} N.~N., 1988, Soviet Astronomy Letters, 14, 334

\bibitem[{{Dall'Ora} et~al.(2014)}]{2014ApJ...787..139D}
{Dall'Ora} M. et~al., 2014, \apj, 787, 139

\bibitem[{{Dessart} \& {Hillier}(2005{\natexlab{a}})}]{2005A&A...439..671D}
{Dessart} L., {Hillier} D.~J., 2005{\natexlab{a}}, \aap, 439, 671

\bibitem[{{Dessart} \& {Hillier}(2005{\natexlab{b}})}]{2005A&A...437..667D}
{Dessart} L., {Hillier} D.~J., 2005{\natexlab{b}}, \aap, 437, 667

\bibitem[{{Dessart} \& {Hillier}(2005{\natexlab{c}})}]{2005ASPC..332..415D}
{Dessart} L., {Hillier} D.~J., 2005{\natexlab{c}}, in R.~{Humphreys},
  K.~{Stanek}, eds, The Fate of the Most Massive Stars. Astronomical Society of
  the Pacific Conference Series, Vol. 332, p. 415

\bibitem[{{Dessart} \& {Hillier}(2006)}]{2006A&A...447..691D}
{Dessart} L., {Hillier} D.~J., 2006, \aap, 447, 691

\bibitem[{{Dessart} et~al.(2008)}]{2008ApJ...675..644D}
{Dessart} L. et~al., 2008, \apj, 675, 644

\bibitem[{{Eastman} et~al.(1996){Eastman}, {Schmidt} \&
  {Kirshner}}]{1996ApJ...466..911E}
{Eastman} R.~G., {Schmidt} B.~P., {Kirshner} R., 1996, \apj, 466, 911

\bibitem[{{Elmhamdi} et~al.(2003{\natexlab{a}}){Elmhamdi}, {Chugai} \&
  {Danziger}}]{2003A&A...404.1077E}
{Elmhamdi} A., {Chugai} N.~N., {Danziger} I.~J., 2003{\natexlab{a}}, \aap, 404,
  1077

\bibitem[{{Elmhamdi} et~al.(2003{\natexlab{b}})}]{2003MNRAS.338..939E}
{Elmhamdi} A. et~al., 2003{\natexlab{b}}, \mnras, 338, 939

\bibitem[{{Fassia} et~al.(2001)}]{2001MNRAS.325..907F}
{Fassia} A. et~al., 2001, \mnras, 325, 907

\bibitem[{{Fisher} et~al.(1997){Fisher}, {Branch}, {Nugent} \&
  {Baron}}]{1997ApJ...481L..89F}
{Fisher} A., {Branch} D., {Nugent} P., {Baron} E., 1997, \apjl, 481, L89

\bibitem[{{Fisher} et~al.(1999){Fisher}, {Branch}, {Hatano} \&
  {Baron}}]{1999MNRAS.304...67F}
{Fisher} A., {Branch} D., {Hatano} K., {Baron} E., 1999, \mnras, 304, 67

\bibitem[{{Hamuy}(2003)}]{2003ApJ...582..905H}
{Hamuy} M., 2003, \apj, 582, 905

\bibitem[{{Hamuy} \& {Suntzeff}(1990)}]{1990AJ.....99.1146H}
{Hamuy} M., {Suntzeff} N.~B., 1990, \aj, 99, 1146

\bibitem[{{Hamuy} et~al.(1994){Hamuy}, {Suntzeff}, {Heathcote}, {Walker},
  {Gigoux} \& {Phillips}}]{1994PASP..106..566H}
{Hamuy} M., {Suntzeff} N.~B., {Heathcote} S.~R., {Walker} A.~R., {Gigoux} P.,
  {Phillips} M.~M., 1994, \pasp, 106, 566

\bibitem[{{Hamuy} et~al.(2001)}]{2001ApJ...558..615H}
{Hamuy} M. et~al., 2001, \apj, 558, 615

\bibitem[{{Hamuy}(2001)}]{2001PhDT.......173H}
{Hamuy} M.~A., 2001, Ph.D. thesis, The University of Arizona

\bibitem[{{Hanuschik} \& {Dachs}(1987)}]{1987A&A...182L..29H}
{Hanuschik} R.~W., {Dachs} J., 1987, \aap, 182, L29

\bibitem[{{Heger} et~al.(2003){Heger}, {Fryer}, {Woosley}, {Langer} \&
  {Hartmann}}]{2003ApJ...591..288H}
{Heger} A., {Fryer} C.~L., {Woosley} S.~E., {Langer} N., {Hartmann} D.~H.,
  2003, \apj, 591, 288

\bibitem[{{Horne}(1986)}]{1986PASP...98..609H}
{Horne} K., 1986, \pasp, 98, 609

\bibitem[{{Inserra} et~al.(2012{\natexlab{a}}){Inserra}, {Baron} \&
  {Turatto}}]{2012MNRAS.422.1178I}
{Inserra} C., {Baron} E., {Turatto} M., 2012{\natexlab{a}}, \mnras, 422, 1178

\bibitem[{{Inserra} et~al.(2011)}]{2011MNRAS.417..261I}
{Inserra} C. et~al., 2011, \mnras, 417, 261

\bibitem[{{Inserra} et~al.(2012{\natexlab{b}})}]{2012MNRAS.422.1122I}
{Inserra} C. et~al., 2012{\natexlab{b}}, \mnras, 422, 1122

\bibitem[{{Inserra} et~al.(2013)}]{2013A&A...555A.142I}
{Inserra} C. et~al., 2013, \aap, 555, A142

\bibitem[{{Jones} et~al.(2009)}]{2009ApJ...696.1176J}
{Jones} M.~I. et~al., 2009, \apj, 696, 1176

\bibitem[{{Kasen} \& {Woosley}(2009)}]{2009ApJ...703.2205K}
{Kasen} D., {Woosley} S.~E., 2009, \apj, 703, 2205

\bibitem[{{Kirshner} \& {Kwan}(1974)}]{1974ApJ...193...27K}
{Kirshner} R.~P., {Kwan} J., 1974, \apj, 193, 27

\bibitem[{{Landolt}(2009)}]{2009AJ....137.4186L}
{Landolt} A.~U., 2009, \aj, 137, 4186

\bibitem[{{Leonard} et~al.(2002{\natexlab{a}})}]{2002AJ....124.2490L}
{Leonard} D.~C. et~al., 2002{\natexlab{a}}, \aj, 124, 2490

\bibitem[{{Leonard} et~al.(2002{\natexlab{b}})}]{2002PASP..114...35L}
{Leonard} D.~C. et~al., 2002{\natexlab{b}}, \pasp, 114, 35

\bibitem[{{Maguire} et~al.(2010)}]{2010MNRAS.404..981M}
{Maguire} K. et~al., 2010, \mnras, 404, 981

\bibitem[{{Makarov} et~al.(2014){Makarov}, {Prugniel}, {Terekhova}, {Courtois}
  \& {Vauglin}}]{2014A&A...570A..13M}
{Makarov} D., {Prugniel} P., {Terekhova} N., {Courtois} H., {Vauglin} I., 2014,
  \aap, 570, A13

\bibitem[{{Marion} et~al.(2014)}]{2014ApJ...781...69M}
{Marion} G.~H. et~al., 2014, \apj, 781, 69

\bibitem[{{Menzies} et~al.(1987)}]{1987MNRAS.227P..39M}
{Menzies} J.~W. et~al., 1987, \mnras, 227, 39P

\bibitem[{{Milisavljevic} et~al.(2013)}]{2013ApJ...767...71M}
{Milisavljevic} D. et~al., 2013, \apj, 767, 71

\bibitem[{{Oke}(1990)}]{1990AJ.....99.1621O}
{Oke} J.~B., 1990, \aj, 99, 1621

\bibitem[{{Olivares} et~al.(2010)}]{2010ApJ...715..833O}
{Olivares} E.~F. et~al., 2010, \apj, 715, 833

\bibitem[{{Pastorello} et~al.(2005)}]{2005MNRAS.360..950P}
{Pastorello} A. et~al., 2005, \mnras, 360, 950

\bibitem[{{Pastorello} et~al.(2009)}]{2009MNRAS.394.2266P}
{Pastorello} A. et~al., 2009, \mnras, 394, 2266

\bibitem[{{Pastorello} et~al.(2012)}]{2012A&A...537A.141P}
{Pastorello} A. et~al., 2012, \aap, 537, A141

\bibitem[{{Pastorello} et~al.(2015)}]{2015MNRAS.449.1921P}
{Pastorello} A. et~al., 2015, \mnras, 449, 1921

\bibitem[{{Patat} et~al.(1994){Patat}, {Barbon}, {Cappellaro} \&
  {Turatto}}]{1994A&A...282..731P}
{Patat} F., {Barbon} R., {Cappellaro} E., {Turatto} M., 1994, \aap, 282, 731

\bibitem[{{Poole} et~al.(2008)}]{2008MNRAS.383..627P}
{Poole} T.~S. et~al., 2008, \mnras, 383, 627

\bibitem[{{Pumo} \& {Zampieri}(2011)}]{2011ApJ...741...41P}
{Pumo} M.~L., {Zampieri} L., 2011, \apj, 741, 41

\bibitem[{{Pumo} \& {Zampieri}(2013)}]{2013MNRAS.434.3445P}
{Pumo} M.~L., {Zampieri} L., 2013, \mnras, 434, 3445

\bibitem[{{Pumo} et~al.(2010){Pumo}, {Zampieri} \&
  {Turatto}}]{2010MSAIS..14..123P}
{Pumo} M.~L., {Zampieri} L., {Turatto} M., 2010, Memorie della Societa
  Astronomica Italiana Supplementi, 14, 123

\bibitem[{{Pumo} et~al.(2009)}]{2009ApJ...705L.138P}
{Pumo} M.~L. et~al., 2009, \apjl, 705, L138

\bibitem[{{Quimby} et~al.(2007){Quimby}, {Wheeler}, {H{\"o}flich}, {Akerlof},
  {Brown} \& {Rykoff}}]{2007ApJ...666.1093Q}
{Quimby} R.~M., {Wheeler} J.~C., {H{\"o}flich} P., {Akerlof} C.~W., {Brown}
  P.~J., {Rykoff} E.~S., 2007, \apj, 666, 1093

\bibitem[{{Rabinak} \& {Waxman}(2011)}]{2011ApJ...728...63R}
{Rabinak} I., {Waxman} E., 2011, \apj, 728, 63

\bibitem[{{Sagar} et~al.(2012){Sagar}, {Kumar}, {Omar} \&
  {Joshi}}]{2012ASInC...4..173S}
{Sagar} R., {Kumar} B., {Omar} A., {Joshi} Y.~C., 2012, in Astronomical Society
  of India Conference Series. Astronomical Society of India Conference Series,
  Vol.~4, p. 173

\bibitem[{{Sahu} et~al.(2006){Sahu}, {Anupama}, {Srividya} \&
  {Muneer}}]{2006MNRAS.372.1315S}
{Sahu} D.~K., {Anupama} G.~C., {Srividya} S., {Muneer} S., 2006, \mnras, 372,
  1315

\bibitem[{{Schlafly} \& {Finkbeiner}(2011)}]{2011ApJ...737..103S}
{Schlafly} E.~F., {Finkbeiner} D.~P., 2011, \apj, 737, 103

\bibitem[{{Shivvers} et~al.(2014){Shivvers}, {Mauerhan}, {Leonard},
  {Filippenko} \& {Fox}}]{2014arXiv1408.1404S}
{Shivvers} I., {Mauerhan} J.~C., {Leonard} D.~C., {Filippenko} A.~V., {Fox}
  O.~D., 2014, ArXiv e-prints

\bibitem[{{Smartt} et~al.(2009){Smartt}, {Eldridge}, {Crockett} \&
  {Maund}}]{2009MNRAS.395.1409S}
{Smartt} S.~J., {Eldridge} J.~J., {Crockett} R.~M., {Maund} J.~R., 2009,
  \mnras, 395, 1409

\bibitem[{{Stetson}(1987)}]{1987PASP...99..191S}
{Stetson} P.~B., 1987, \pasp, 99, 191

\bibitem[{{Tak{\'a}ts} \& {Vink{\'o}}(2012)}]{2012MNRAS.419.2783T}
{Tak{\'a}ts} K., {Vink{\'o}} J., 2012, \mnras, 419, 2783

\bibitem[{{Tak{\'a}ts} et~al.(2014)}]{2014MNRAS.438..368T}
{Tak{\'a}ts} K. et~al., 2014, \mnras, 438, 368

\bibitem[{{Theureau} et~al.(2007){Theureau}, {Hanski}, {Coudreau}, {Hallet} \&
  {Martin}}]{2007A&A...465...71T}
{Theureau} G., {Hanski} M.~O., {Coudreau} N., {Hallet} N., {Martin} J.~M.,
  2007, \aap, 465, 71

\bibitem[{{Tomasella} et~al.(2013)}]{2013MNRAS.434.1636T}
{Tomasella} L. et~al., 2013, \mnras, 434, 1636

\bibitem[{{Tsvetkov} et~al.(2008){Tsvetkov}, {Goranskij} \&
  {Pavlyuk}}]{2008PZ.....28....8T}
{Tsvetkov} D.~Y., {Goranskij} V., {Pavlyuk} N., 2008, Peremennye Zvezdy, 28, 8

\bibitem[{{Tully} et~al.(2009){Tully}, {Rizzi}, {Shaya}, {Courtois}, {Makarov}
  \& {Jacobs}}]{2009AJ....138..323T}
{Tully} R.~B., {Rizzi} L., {Shaya} E.~J., {Courtois} H.~M., {Makarov} D.~I.,
  {Jacobs} B.~A., 2009, \aj, 138, 323

\bibitem[{{Turatto} et~al.(2003){Turatto}, {Benetti} \&
  {Cappellaro}}]{2003fthp.conf..200T}
{Turatto} M., {Benetti} S., {Cappellaro} E., 2003, in W.~{Hillebrandt},
  B.~{Leibundgut}, eds, From Twilight to Highlight: The Physics of Supernovae.
  p. 200

\bibitem[{{Utrobin} \& {Chugai}(2009)}]{2009A&A...506..829U}
{Utrobin} V.~P., {Chugai} N.~N., 2009, \aap, 506, 829

\bibitem[{{Valenti} et~al.(2015)}]{2015MNRAS.448.2608V}
{Valenti} S. et~al., 2015, \mnras, 448, 2608

\bibitem[{{van Dokkum}(2001)}]{2001PASP..113.1420V}
{van Dokkum} P.~G., 2001, \pasp, 113, 1420

\bibitem[{{Zampieri} et~al.(2003)}]{2003MNRAS.338..711Z}
{Zampieri} L., {Pastorello} A., {Turatto} M., {Cappellaro} E., {Benetti} S.,
  {Altavilla} G., {Mazzali} P., {Hamuy} M., 2003, \mnras, 338, 711

\bibitem[{{Zheng} et~al.(2013){Zheng}, {Blanchard}, {Cenko}, {Filippenko} \&
  {Cucchiara}}]{2013ATel.4823....1Z}
{Zheng} W., {Blanchard} P., {Cenko} S.~B., {Filippenko} A.~V., {Cucchiara} A.,
  2013, The Astronomer's Telegram, 4823, 1

\end{thebibliography}

\label{lastpage}

\appendix
\section{Photometry}
\label{app:photometry}
 Photometric images were bias subtracted and flat fielded, then cosmic ray removal were performed using standard tasks
 available in {\iraf}\footnote{{\iraf} stands for
 Image Reduction and Analysis Facility distributed by the National Optical Astronomy Observatories which
 is operated by the Association of Universities for research in Astronomy,
 Inc. under co-operative agreement with the National Science Foundation.}.
 Due to the position of SN in its host galaxy, we performed PSF photometry on all frames using {\daophot}\,\footnote{ {\daophot} stands for Dominion Astrophysical Observatory Photometry \citep{1987PASP...99..191S}.} routines. The measured SN flux was significantly affected by the host galaxy flux,
 therefore an annulus has been chosen conservatively to estimate the host galaxy background.

To calibrate the instrumental light curves of \sn\ in \ubvri, obtained from ARIES telescopes, four \citet{2009AJ....137.4186L} standard fields PG~1323, PG~1633, SA~104 and SA~107 were observed on 8 March, 2014 with 104-cm ST under good photometric sky condition with low atmospheric extinction and a typical image FWHM $ \sim$1\arcsec.9 in \textit{V} band. Multiple standard fields at different airmasses were used to compute the zero points, colour coefficients as well as extinction for different filters. In total 27 standard stars were used for calibration having \textit{V} magnitudes from 11.0 to 15.4. The root-mean-square scatter between standard and the re-calibrated magnitudes of these Landolt stars were found to be $ \sim0.4 $ mag in \textit{U}, $ \sim0.2 $ mag in \textit{B}, $ \sim0.1 $ mag in \textit{V,R} and $ \sim0.3 $ mag in \textit{I} band.
We used the obtained coefficients to standardize 11 non-variable and isolated field stars in the SN frame, which are tabulated in Table.~\ref{tab:photstar} . These secondary standards allows us to calibrate the SN light curve obtained using differential photometry. Similarly to standardize the data collected from LCOGT in \textit{BVgri} bands, several secondary standard in the SN field are selected, few of which are also listed in Table.~\ref{tab:photstar}. Since, \textit{B} and \textit{V} bands are in common to ARIES and LOCGT data, to have an uniformity we preferred to used LCOGT secondary standards for all \textit{BV} bands data.
\swift/UVOT data were analyzed following the methods
described by \cite{2008MNRAS.383..627P} and \cite{2009AJ....137.4517B} but adopting
the revised zeropoints and sensitivity from \cite{2011AIPC.1358..373B}.
Template images in all UVOT bands are obtained on 13 March, 2015 (755d), which are used to estimate background fluxes, using same aperture as used for SN measurement. These fluxes are subtracted from SN to eliminate flux contamination due to host galaxy.
The final photometry in the \ubvri, \textit{gri} and UVOT bands is tabulated in Table~\ref{tab:photsn}.

\begin{table*}
  \caption{Coordinates ($\alpha, \delta$) and calibrated
           magnitudes of secondary standard stars in the field of \sn. The quoted errors in magnitude include both photometric and calibration
           errors and it denote 1$\sigma$ uncertainty.}
  \label{tab:photstar}
Secondary standards used to calibrate \textit{URI} data.\\
  \begin{tabular}{cccccc}
     \hline
     $\alpha_{\rm J2000}$& $\delta_{\rm J2000}$& $U$                 &  $R$                &  $I$     \\
         (h m s)&(\degr\,\arcmin\,\arcsec)     &(mag)                &(mag)                &(mag)     \\
     \hline
14:32:38 & +9:56:39                            & 17.654  $\pm$ 0.085 & 15.452 $\pm$  0.016 & 14.967  $\pm$ 0.014  \\
14:32:36 & +9:56:00                            & 17.406  $\pm$ 0.028 & 16.134 $\pm$  0.017 & 15.744  $\pm$ 0.016  \\
14:32:33 & +9:54:45                            & 16.968  $\pm$ 0.022 & 15.851 $\pm$  0.017 & 15.440  $\pm$ 0.016  \\
14:32:38 & +9:52:24                            & 17.709  $\pm$ 0.034 & 16.034 $\pm$  0.020 & 15.550  $\pm$ 0.021  \\
14:32:28 & +9:49:43                            & 19.362  $\pm$ 0.204 & 15.628 $\pm$  0.015 & 14.616  $\pm$ 0.015  \\
14:32:45 & +9:48:37                            & 18.118  $\pm$ 0.062 & 16.549 $\pm$  0.017 & 16.109  $\pm$ 0.020  \\
14:33:02 & +9:54:41                            & 18.789  $\pm$ 0.097 & 15.133 $\pm$  0.013 & 14.096  $\pm$ 0.015  \\
14:33:07 & +9:55:52                            & 17.354  $\pm$ 0.023 & 16.085 $\pm$  0.013 & 15.649  $\pm$ 0.013  \\
14:33:02 & +9:56:03                            & 17.828  $\pm$ 0.031 & 16.422 $\pm$  0.012 & 15.996  $\pm$ 0.014  \\
14:33:01 & +9:57:25                            & 14.520  $\pm$ 0.011 & 13.770 $\pm$  0.009 & 13.418  $\pm$ 0.013  \\
14:32:46 & +9:56:10                            & 17.918  $\pm$ 0.071 & 16.882 $\pm$  0.013 & 16.469  $\pm$ 0.015  \\
     \hline
  \end{tabular}
 \\ Secondary standards used to calibrate \textit{BVgri} data.\\
    \begin{tabular}{cccccccc}
     \hline
     $\alpha_{\rm J2000}$&       $\delta_{\rm J2000}$& $B$ & $V$ &  $g$&  $r$&  $i$\\
              (h m s)&(\degr\,\arcmin\,\arcsec)      &(mag)&(mag)&(mag)&(mag)&(mag)\\
     \hline
14:32:24 & +9:52:12 & 18.947 $\pm$ 0.098 & 18.518 $\pm$ 0.032 & 18.749 $\pm$  0.019 & 18.336 $\pm$ 0.022 & 18.241 $\pm$  0.026 \\
14:32:26 & +9:50:09 & 16.924 $\pm$ 0.013 & 16.374 $\pm$ 0.011 & 16.648 $\pm$  0.011 & 16.147 $\pm$ 0.010 & 16.009 $\pm$  0.026 \\
14:32:32 & +9:47:16 & 17.557 $\pm$ 0.040 & 16.977 $\pm$ 0.025 & 17.321 $\pm$  0.025 & 16.788 $\pm$ 0.016 & 16.629 $\pm$  0.021 \\
14:32:32 & +9:57:48 & 17.994 $\pm$ 0.021 & 17.428 $\pm$ 0.016 & 17.733 $\pm$  0.013 & 17.196 $\pm$ 0.012 & 17.055 $\pm$  0.022 \\
14:32:37 & +9:58:48 & 18.937 $\pm$ 0.048 & 17.948 $\pm$ 0.023 & 18.461 $\pm$  0.019 & 17.413 $\pm$ 0.013 & 16.961 $\pm$  0.013 \\
14:32:40 & +9:45:54 & 14.499 $\pm$ 0.016 & 14.064 $\pm$ 0.010 & 14.286 $\pm$  0.010 & 13.912 $\pm$ 0.010 & 13.814 $\pm$  0.011 \\
14:32:44 & +9:54:24 & 16.593 $\pm$ 0.012 & 15.759 $\pm$ 0.010 & 16.166 $\pm$  0.011 & 15.426 $\pm$ 0.016 & 15.203 $\pm$  0.023 \\
14:32:48 & +9:46:33 & 19.776 $\pm$ 0.083 & 19.286 $\pm$ 0.056 & 19.547 $\pm$  0.057 & 19.189 $\pm$ 0.042 & 19.132 $\pm$  0.173 \\
14:32:59 & +9:45:23 & 17.241 $\pm$ 0.021 & 16.804 $\pm$ 0.013 & 17.022 $\pm$  0.013 & 16.635 $\pm$ 0.012 & 16.497 $\pm$  0.013 \\
14:33:02 & +9:54:41 & 17.423 $\pm$ 0.016 & 16.078 $\pm$ 0.011 & 16.746 $\pm$  0.011 & 15.450 $\pm$ 0.011 & 14.661 $\pm$  0.010 \\
14:33:07 & +9:55:52 & 17.096 $\pm$ 0.014 & 16.462 $\pm$ 0.012 & 16.746 $\pm$  0.011 & 16.216 $\pm$ 0.011 & 16.033 $\pm$  0.011 \\
14:33:11 & +9:59:27 & 17.485 $\pm$ 0.015 & 16.674 $\pm$ 0.012 & 17.092 $\pm$  0.012 & 16.309 $\pm$ 0.011 & 16.042 $\pm$  0.012 \\
14:33:13 & +9:49:03 & 16.591 $\pm$ 0.012 & 16.059 $\pm$ 0.011 & 16.329 $\pm$  0.011 & 15.865 $\pm$ 0.011 & 15.725 $\pm$  0.011 \\
14:33:14 & +9:52:02 & 18.502 $\pm$ 0.060 & 17.726 $\pm$ 0.028 & 18.161 $\pm$  0.016 & 17.349 $\pm$ 0.012 & 17.049 $\pm$  0.016 \\
14:33:15 & +9:51:37 & 15.060 $\pm$ 0.012 & 14.534 $\pm$ 0.010 & 14.776 $\pm$  0.010 & 14.344 $\pm$ 0.010 & 14.230 $\pm$  0.011 \\
14:33:18 & +9:47:11 & 17.575 $\pm$ 0.168 & 17.052 $\pm$ 0.062 & 17.586 $\pm$  0.068 & 16.800 $\pm$ 0.013 & 16.531 $\pm$  0.014 \\
     \hline
  \end{tabular}

\end{table*}

\onecolumn
\begin{landscape}
\subsection{Photometric data} \label{app.data}
  \fontsize{2.6mm}{2.8mm}\selectfont
  \begin{longtable}
  {c c r c c c c c c c c l}
    \caption{Photometric evolution of \sn. Errors denote $1\sigma$ uncertainty.}
    \label{tab:photsn}\\
  \hline
  UT Date&JD&Phase$^{a}$&$U$&$B$&$V$&$R$&$I$&$g$&$r$&$i$&Tel$^{b}$ \\
  (yyyy/mm/dd)&2456000+&(day)&(mag)&(mag)&(mag)&(mag)&(mag)&(mag)&(mag)&(mag)& /Inst\\
  \hline

2013-02-15.50  &  339.00   &   -1.00   &           ---        &          ---        &          ---        &          ---        &          ---        &          ---        &         $ > $18.500      &          ---        &    6  \\
2013-02-17.50  &  341.00   &    1.00   &           ---        &          ---        &          ---        &          ---        &          ---        &          ---        &  17.600 $\pm$ 0.095 &          ---        &    6  \\
2013-02-19.20  &  342.70   &    2.70   &           ---        &          ---        &          ---        &          ---        &          ---        &          ---        &  15.093 $\pm$ 0.095 &          ---        &    5  \\
2013-02-20.29  &  343.79   &    3.79   &           ---        &          ---        &          ---        &          ---        &          ---        &  15.007 $\pm$ 0.022 &  15.094 $\pm$ 0.047 &  15.297 $\pm$ 0.010 &  2,5  \\
2013-02-20.93  &  344.43   &    4.43   &   14.231 $\pm$ 0.040 &          ---        &          ---        &  14.873 $\pm$ 0.015 &  14.767 $\pm$ 0.030 &          ---        &          ---        &          ---        &    1  \\
2013-02-21.34  &  344.84   &    4.84   &           ---        &  14.812 $\pm$ 0.015 &  14.779 $\pm$ 0.012 &          ---        &          ---        &  14.805 $\pm$ 0.018 &  14.874 $\pm$ 0.011 &  14.956 $\pm$ 0.017 &    5  \\
2013-02-21.57  &  345.07   &    5.07   &           ---        &          ---        &          ---        &          ---        &          ---        &  14.964 $\pm$ 0.016 &  14.767 $\pm$ 0.068 &  14.994 $\pm$ 0.064 &    3  \\
2013-02-22.34  &  345.84   &    5.84   &           ---        &  14.813 $\pm$ 0.015 &  14.747 $\pm$ 0.018 &          ---        &          ---        &  14.825 $\pm$ 0.018 &  14.759 $\pm$ 0.017 &  14.855 $\pm$ 0.013 &    5  \\
2013-02-22.90  &  346.40   &    6.40   &   14.173 $\pm$ 0.040 &  14.804 $\pm$ 0.025 &          ---        &  14.730 $\pm$ 0.014 &  14.586 $\pm$ 0.029 &          ---        &          ---        &          ---        &    1  \\
2013-02-23.34  &  346.84   &    6.84   &           ---        &  14.793 $\pm$ 0.013 &  14.705 $\pm$ 0.023 &          ---        &          ---        &  14.734 $\pm$ 0.028 &  14.744 $\pm$ 0.037 &  14.740 $\pm$ 0.030 &    5  \\
2013-02-24.34  &  347.84   &    7.84   &           ---        &  14.793 $\pm$ 0.014 &          ---        &          ---        &          ---        &  14.761 $\pm$ 0.013 &  14.699 $\pm$ 0.010 &  14.763 $\pm$ 0.012 &    5  \\
2013-02-24.80  &  348.30   &    8.30   &   14.231 $\pm$ 0.041 &  14.868 $\pm$ 0.025 &          ---        &  14.677 $\pm$ 0.014 &  14.517 $\pm$ 0.030 &          ---        &          ---        &          ---        &    1  \\
2013-02-25.34  &  348.84   &    8.84   &           ---        &  14.838 $\pm$ 0.015 &  14.691 $\pm$ 0.019 &          ---        &          ---        &  14.776 $\pm$ 0.013 &  14.721 $\pm$ 0.025 &  14.776 $\pm$ 0.016 &    5  \\
2013-02-26.34  &  349.84   &    9.84   &           ---        &  14.871 $\pm$ 0.011 &  14.729 $\pm$ 0.021 &          ---        &          ---        &  14.751 $\pm$ 0.020 &  14.717 $\pm$ 0.021 &  14.786 $\pm$ 0.009 &    5  \\
2013-02-26.90  &  350.40   &   10.40   &   14.342 $\pm$ 0.041 &  14.885 $\pm$ 0.025 &          ---        &  14.698 $\pm$ 0.014 &  14.552 $\pm$ 0.030 &          ---        &          ---        &          ---        &    1  \\
2013-02-27.34  &  350.84   &   10.84   &           ---        &  14.880 $\pm$ 0.014 &  14.783 $\pm$ 0.011 &          ---        &          ---        &  14.903 $\pm$ 0.013 &  14.685 $\pm$ 0.022 &  14.871 $\pm$ 0.021 &    5  \\
2013-02-27.80  &  351.30   &   11.30   &   14.402 $\pm$ 0.041 &  14.916 $\pm$ 0.025 &          ---        &  14.712 $\pm$ 0.014 &  14.580 $\pm$ 0.029 &          ---        &          ---        &          ---        &    1  \\
2013-02-28.34  &  351.84   &   11.84   &           ---        &  14.860 $\pm$ 0.015 &  14.814 $\pm$ 0.018 &          ---        &          ---        &  14.831 $\pm$ 0.016 &  14.741 $\pm$ 0.014 &  14.822 $\pm$ 0.015 &    5  \\
2013-02-28.86  &  352.36   &   12.36   &   14.497 $\pm$ 0.040 &  14.974 $\pm$ 0.025 &          ---        &  14.733 $\pm$ 0.014 &  14.612 $\pm$ 0.029 &          ---        &          ---        &          ---        &    1  \\
2013-03-01.44  &  352.94   &   12.94   &           ---        &  15.018 $\pm$ 0.005 &  14.901 $\pm$ 0.006 &          ---        &          ---        &  14.935 $\pm$ 0.005 &  14.809 $\pm$ 0.005 &  14.911 $\pm$ 0.009 &    5  \\
2013-03-02.44  &  353.94   &   13.94   &           ---        &  14.950 $\pm$ 0.006 &  14.889 $\pm$ 0.010 &          ---        &          ---        &  14.925 $\pm$ 0.007 &  14.777 $\pm$ 0.009 &  14.934 $\pm$ 0.009 &    5  \\
2013-03-03.39  &  354.89   &   14.89   &           ---        &  15.092 $\pm$ 0.017 &  14.826 $\pm$ 0.018 &          ---        &          ---        &  14.917 $\pm$ 0.003 &  14.798 $\pm$ 0.006 &  14.951 $\pm$ 0.007 &    5  \\
2013-03-04.44  &  355.94   &   15.94   &           ---        &  15.046 $\pm$ 0.005 &  14.920 $\pm$ 0.005 &          ---        &          ---        &  14.957 $\pm$ 0.005 &  14.802 $\pm$ 0.005 &  14.945 $\pm$ 0.009 &    5  \\
2013-03-04.84  &  356.34   &   16.34   &   14.715 $\pm$ 0.040 &  15.073 $\pm$ 0.025 &          ---        &  14.770 $\pm$ 0.014 &  14.668 $\pm$ 0.029 &          ---        &          ---        &          ---        &    1  \\
2013-03-05.32  &  356.82   &   16.82   &           ---        &  15.102 $\pm$ 0.023 &  14.885 $\pm$ 0.018 &          ---        &          ---        &  14.972 $\pm$ 0.015 &  14.753 $\pm$ 0.024 &  14.893 $\pm$ 0.027 &    5  \\
2013-03-06.32  &  357.82   &   17.82   &           ---        &  15.122 $\pm$ 0.012 &  14.905 $\pm$ 0.021 &          ---        &          ---        &  15.007 $\pm$ 0.019 &  14.750 $\pm$ 0.026 &  14.930 $\pm$ 0.023 &    5  \\
2013-03-06.85  &  358.35   &   18.35   &   14.879 $\pm$ 0.040 &  15.124 $\pm$ 0.025 &          ---        &  14.768 $\pm$ 0.014 &  14.652 $\pm$ 0.029 &          ---        &          ---        &          ---        &    1  \\
2013-03-07.82  &  359.32   &   19.32   &   14.951 $\pm$ 0.040 &  15.162 $\pm$ 0.025 &          ---        &  14.762 $\pm$ 0.014 &  14.642 $\pm$ 0.029 &          ---        &          ---        &          ---        &    1  \\
2013-03-08.31  &  359.81   &   19.81   &           ---        &  15.198 $\pm$ 0.017 &          ---        &          ---        &          ---        &          ---        &          ---        &          ---        &    5  \\
2013-03-09.36  &  360.86   &   20.86   &           ---        &  15.196 $\pm$ 0.016 &  14.886 $\pm$ 0.012 &          ---        &          ---        &  15.026 $\pm$ 0.009 &  14.783 $\pm$ 0.009 &          ---        &    5  \\
2013-03-10.35  &  361.85   &   21.85   &           ---        &  15.254 $\pm$ 0.016 &  14.935 $\pm$ 0.026 &          ---        &          ---        &  15.075 $\pm$ 0.012 &  14.832 $\pm$ 0.013 &  14.934 $\pm$ 0.015 &    5  \\
2013-03-11.33  &  362.83   &   22.83   &           ---        &  15.236 $\pm$ 0.010 &  14.963 $\pm$ 0.014 &          ---        &          ---        &  15.067 $\pm$ 0.016 &  14.814 $\pm$ 0.019 &  14.939 $\pm$ 0.021 &    5  \\
2013-03-12.37  &  363.87   &   23.87   &           ---        &  15.363 $\pm$ 0.016 &  14.952 $\pm$ 0.019 &          ---        &          ---        &  15.091 $\pm$ 0.014 &  14.827 $\pm$ 0.004 &  14.923 $\pm$ 0.004 &    5  \\
2013-03-12.80  &  364.30   &   24.30   &   15.503 $\pm$ 0.041 &  15.354 $\pm$ 0.025 &  14.994 $\pm$ 0.012 &  14.756 $\pm$ 0.014 &  14.605 $\pm$ 0.029 &          ---        &          ---        &          ---        &    1  \\
2013-03-14.44  &  365.94   &   25.94   &           ---        &  15.422 $\pm$ 0.012 &  15.036 $\pm$ 0.013 &          ---        &          ---        &  15.202 $\pm$ 0.011 &  14.890 $\pm$ 0.015 &  14.985 $\pm$ 0.018 &    5  \\
2013-03-15.88  &  367.38   &   27.38   &   15.819 $\pm$ 0.042 &  15.486 $\pm$ 0.025 &  15.056 $\pm$ 0.012 &  14.793 $\pm$ 0.014 &  14.636 $\pm$ 0.029 &          ---        &          ---        &          ---        &    1  \\
2013-03-19.94  &  371.44   &   31.44   &   16.145 $\pm$ 0.044 &  15.665 $\pm$ 0.025 &  15.166 $\pm$ 0.012 &  14.867 $\pm$ 0.014 &  14.693 $\pm$ 0.029 &          ---        &          ---        &          ---        &    1  \\
2013-03-20.44  &  371.94   &   31.94   &           ---        &  15.695 $\pm$ 0.007 &  15.113 $\pm$ 0.005 &          ---        &          ---        &  15.383 $\pm$ 0.003 &  14.934 $\pm$ 0.004 &  15.035 $\pm$ 0.007 &    5  \\
2013-03-21.71  &  373.21   &   33.21   &   16.243 $\pm$ 0.053 &          ---        &  15.203 $\pm$ 0.013 &  14.900 $\pm$ 0.015 &  14.692 $\pm$ 0.031 &          ---        &          ---        &          ---        &    1  \\
2013-03-24.43  &  375.93   &   35.93   &           ---        &  15.812 $\pm$ 0.010 &  15.230 $\pm$ 0.007 &          ---        &          ---        &  15.464 $\pm$ 0.005 &  15.003 $\pm$ 0.005 &          ---        &    5  \\
2013-03-24.90  &  376.40   &   36.40   &           ---        &  15.848 $\pm$ 0.027 &  15.248 $\pm$ 0.013 &  14.927 $\pm$ 0.014 &          ---        &          ---        &          ---        &          ---        &    1  \\
2013-03-25.75  &  377.25   &   37.25   &   16.633 $\pm$ 0.059 &  15.866 $\pm$ 0.027 &  15.257 $\pm$ 0.009 &  14.936 $\pm$ 0.015 &  14.718 $\pm$ 0.030 &          ---        &          ---        &          ---        &    1  \\
2013-03-26.38  &  377.88   &   37.88   &           ---        &  15.911 $\pm$ 0.012 &  15.245 $\pm$ 0.012 &          ---        &          ---        &          ---        &          ---        &          ---        &    5  \\
2013-03-28.81  &  380.31   &   40.31   &   16.681 $\pm$ 0.062 &  15.968 $\pm$ 0.028 &  15.282 $\pm$ 0.013 &  14.966 $\pm$ 0.015 &  14.729 $\pm$ 0.030 &          ---        &          ---        &          ---        &    1  \\
2013-03-29.78  &  381.28   &   41.28   &   16.814 $\pm$ 0.104 &  15.982 $\pm$ 0.029 &  15.303 $\pm$ 0.013 &  14.971 $\pm$ 0.014 &  14.722 $\pm$ 0.030 &          ---        &          ---        &          ---        &    1  \\
2013-03-30.87  &  382.37   &   42.37   &   16.910 $\pm$ 0.066 &  16.035 $\pm$ 0.027 &  15.316 $\pm$ 0.013 &  14.997 $\pm$ 0.015 &  14.763 $\pm$ 0.029 &          ---        &          ---        &          ---        &    1  \\
2013-04-01.38  &  383.88   &   43.88   &           ---        &  16.068 $\pm$ 0.013 &  15.278 $\pm$ 0.014 &          ---        &          ---        &  15.619 $\pm$ 0.014 &  15.073 $\pm$ 0.017 &  15.083 $\pm$ 0.018 &    5  \\
2013-04-05.39  &  387.89   &   47.89   &           ---        &  16.073 $\pm$ 0.204 &  15.277 $\pm$ 0.025 &          ---        &          ---        &  15.640 $\pm$ 0.020 &  15.064 $\pm$ 0.026 &  15.074 $\pm$ 0.028 &    5  \\
2013-04-05.83  &  388.33   &   48.33   &   17.143 $\pm$ 0.055 &  16.074 $\pm$ 0.026 &  15.307 $\pm$ 0.012 &  14.986 $\pm$ 0.014 &  14.736 $\pm$ 0.029 &          ---        &          ---        &          ---        &    1  \\
2013-04-06.40  &  388.90   &   48.90   &           ---        &  16.183 $\pm$ 0.030 &          ---        &          ---        &          ---        &          ---        &          ---        &          ---        &    5  \\
2013-04-06.85  &  389.35   &   49.35   &   17.200 $\pm$ 0.042 &  16.087 $\pm$ 0.018 &  15.317 $\pm$ 0.008 &  15.117 $\pm$ 0.010 &  14.731 $\pm$ 0.021 &          ---        &          ---        &          ---        &    1  \\
2013-04-07.39  &  389.89   &   49.89   &           ---        &  16.203 $\pm$ 0.010 &  15.340 $\pm$ 0.012 &          ---        &          ---        &  15.693 $\pm$ 0.013 &  15.110 $\pm$ 0.014 &  15.119 $\pm$ 0.012 &    5  \\
2013-04-10.88  &  393.38   &   53.38   &   17.429 $\pm$ 0.047 &  16.243 $\pm$ 0.026 &  15.381 $\pm$ 0.012 &  15.032 $\pm$ 0.014 &  14.751 $\pm$ 0.029 &          ---        &          ---        &          ---        &    1  \\
2013-04-11.85  &  394.35   &   54.35   &   17.487 $\pm$ 0.073 &  16.243 $\pm$ 0.026 &  15.383 $\pm$ 0.012 &  15.030 $\pm$ 0.014 &  14.748 $\pm$ 0.029 &          ---        &          ---        &          ---        &    1  \\
2013-04-14.36  &  396.86   &   56.86   &           ---        &  16.362 $\pm$ 0.011 &  15.372 $\pm$ 0.010 &          ---        &          ---        &  15.771 $\pm$ 0.008 &  15.167 $\pm$ 0.008 &  15.194 $\pm$ 0.009 &    5  \\
2013-04-14.77  &  397.27   &   57.27   &   17.522 $\pm$ 0.052 &  16.288 $\pm$ 0.026 &  15.403 $\pm$ 0.012 &  15.042 $\pm$ 0.014 &  14.761 $\pm$ 0.029 &          ---        &          ---        &          ---        &    1  \\
  \hline
  UT Date&JD&Phase$^{a}$&$U$&$B$&$V$&$R$&$I$&$g$&$r$&$i$&Tel$^{b}$ \\
  (yyyy/mm/dd)&2456000+&(day)&(mag)&(mag)&(mag)&(mag)&(mag)&(mag)&(mag)&(mag)& /Inst\\
  \hline
2013-04-19.32  &  401.82   &   61.82   &           ---        &  16.402 $\pm$ 0.014 &  15.373 $\pm$ 0.011 &          ---        &          ---        &  15.838 $\pm$ 0.007 &  15.197 $\pm$ 0.010 &  15.122 $\pm$ 0.013 &    5  \\
2013-04-20.32  &  402.82   &   62.82   &           ---        &  16.390 $\pm$ 0.012 &  15.420 $\pm$ 0.010 &          ---        &          ---        &  15.804 $\pm$ 0.006 &  15.229 $\pm$ 0.009 &  15.191 $\pm$ 0.012 &    5  \\
2013-04-21.32  &  403.82   &   63.82   &           ---        &  16.444 $\pm$ 0.018 &  15.387 $\pm$ 0.014 &          ---        &          ---        &  15.793 $\pm$ 0.013 &  15.153 $\pm$ 0.018 &  15.150 $\pm$ 0.016 &    5  \\
2013-04-22.77  &  405.27   &   65.27   &   17.637 $\pm$ 0.098 &  16.415 $\pm$ 0.022 &  15.430 $\pm$ 0.009 &  15.061 $\pm$ 0.016 &  14.787 $\pm$ 0.030 &          ---        &          ---        &          ---        &    1  \\
2013-04-23.32  &  405.82   &   65.82   &           ---        &  16.330 $\pm$ 0.026 &  15.411 $\pm$ 0.012 &          ---        &          ---        &  15.820 $\pm$ 0.014 &  15.182 $\pm$ 0.019 &  15.143 $\pm$ 0.017 &    5  \\
2013-04-23.80  &  406.30   &   66.30   &           ---        &          ---        &  15.422 $\pm$ 0.017 &  15.084 $\pm$ 0.015 &  14.781 $\pm$ 0.030 &          ---        &          ---        &          ---        &    1  \\
2013-04-26.28  &  408.78   &   68.78   &           ---        &          ---        &          ---        &          ---        &          ---        &  15.874 $\pm$ 0.018 &  15.099 $\pm$ 0.016 &          ---        &    5  \\
2013-04-28.28  &  410.78   &   70.78   &           ---        &  16.480 $\pm$ 0.019 &  15.406 $\pm$ 0.014 &          ---        &          ---        &  15.818 $\pm$ 0.011 &  15.129 $\pm$ 0.016 &  15.137 $\pm$ 0.016 &    5  \\
2013-04-29.28  &  411.78   &   71.78   &           ---        &  16.520 $\pm$ 0.014 &  15.437 $\pm$ 0.011 &          ---        &          ---        &  15.858 $\pm$ 0.009 &  15.187 $\pm$ 0.008 &  15.175 $\pm$ 0.012 &    5  \\
2013-04-30.28  &  412.78   &   72.78   &           ---        &  16.557 $\pm$ 0.013 &  15.486 $\pm$ 0.012 &          ---        &          ---        &  15.889 $\pm$ 0.011 &  15.227 $\pm$ 0.010 &  15.246 $\pm$ 0.013 &    5  \\
2013-05-01.77  &  414.27   &   74.27   &   17.740 $\pm$ 0.053 &          ---        &  15.444 $\pm$ 0.013 &  15.086 $\pm$ 0.014 &  14.792 $\pm$ 0.030 &          ---        &          ---        &          ---        &    1  \\
2013-05-03.27  &  415.77   &   75.77   &           ---        &  16.617 $\pm$ 0.025 &  15.498 $\pm$ 0.015 &          ---        &          ---        &  15.885 $\pm$ 0.011 &  15.231 $\pm$ 0.012 &          ---        &    5  \\
2013-05-05.75  &  418.25   &   78.25   &   17.841 $\pm$ 0.094 &          ---        &  15.466 $\pm$ 0.012 &  15.107 $\pm$ 0.014 &  14.820 $\pm$ 0.021 &          ---        &          ---        &          ---        &    1  \\
2013-05-06.76  &  419.26   &   79.26   &   17.920 $\pm$ 0.086 &          ---        &  15.457 $\pm$ 0.012 &  15.094 $\pm$ 0.014 &  14.802 $\pm$ 0.029 &          ---        &          ---        &          ---        &    1  \\
2013-05-08.27  &  420.77   &   80.77   &           ---        &  16.620 $\pm$ 0.027 &  15.562 $\pm$ 0.019 &          ---        &          ---        &  15.985 $\pm$ 0.015 &  15.298 $\pm$ 0.013 &  15.224 $\pm$ 0.021 &    5  \\
2013-05-08.79  &  421.29   &   81.29   &   18.113 $\pm$ 0.061 &          ---        &  15.494 $\pm$ 0.012 &  15.123 $\pm$ 0.015 &  14.815 $\pm$ 0.030 &          ---        &          ---        &          ---        &    1  \\
2013-05-09.72  &  422.22   &   82.22   &   18.070 $\pm$ 0.057 &          ---        &  15.510 $\pm$ 0.009 &  15.119 $\pm$ 0.015 &  14.823 $\pm$ 0.030 &          ---        &          ---        &          ---        &    1  \\
2013-05-12.74  &  425.24   &   85.24   &   18.069 $\pm$ 0.078 &          ---        &  15.567 $\pm$ 0.013 &  15.174 $\pm$ 0.015 &  14.815 $\pm$ 0.032 &          ---        &          ---        &          ---        &    1  \\
2013-05-13.24  &  425.74   &   85.74   &           ---        &  16.793 $\pm$ 0.023 &  15.621 $\pm$ 0.015 &          ---        &          ---        &  16.084 $\pm$ 0.010 &  15.360 $\pm$ 0.011 &  15.329 $\pm$ 0.014 &    5  \\
2013-05-15.24  &  427.74   &   87.74   &           ---        &  16.874 $\pm$ 0.021 &  15.653 $\pm$ 0.014 &          ---        &          ---        &  16.127 $\pm$ 0.010 &  15.360 $\pm$ 0.009 &  15.349 $\pm$ 0.013 &    5  \\
2013-05-16.26  &  428.76   &   88.76   &           ---        &  16.827 $\pm$ 0.021 &  15.597 $\pm$ 0.013 &          ---        &          ---        &  16.093 $\pm$ 0.015 &  15.337 $\pm$ 0.012 &          ---        &    5  \\
2013-05-17.74  &  430.24   &   90.24   &   18.634 $\pm$ 0.147 &          ---        &  15.668 $\pm$ 0.013 &  15.223 $\pm$ 0.014 &  14.926 $\pm$ 0.030 &          ---        &          ---        &          ---        &    1  \\
2013-05-18.71  &  431.21   &   91.21   &           ---        &          ---        &  15.718 $\pm$ 0.010 &  15.279 $\pm$ 0.015 &  14.969 $\pm$ 0.031 &          ---        &          ---        &          ---        &    1  \\
2013-05-19.69  &  432.19   &   92.19   &           ---        &          ---        &  15.740 $\pm$ 0.015 &  15.298 $\pm$ 0.016 &  14.978 $\pm$ 0.031 &          ---        &          ---        &          ---        &    1  \\
2013-05-20.26  &  432.76   &   92.76   &           ---        &  17.002 $\pm$ 0.021 &  15.764 $\pm$ 0.015 &          ---        &          ---        &  16.256 $\pm$ 0.021 &  15.474 $\pm$ 0.027 &  15.425 $\pm$ 0.025 &    5  \\
2013-05-21.72  &  434.22   &   94.22   &           ---        &          ---        &  15.783 $\pm$ 0.015 &  15.350 $\pm$ 0.012 &  15.025 $\pm$ 0.034 &          ---        &          ---        &          ---        &    1  \\
2013-05-22.23  &  434.73   &   94.73   &           ---        &  17.154 $\pm$ 0.034 &  15.864 $\pm$ 0.014 &          ---        &          ---        &  16.382 $\pm$ 0.013 &  15.523 $\pm$ 0.016 &  15.466 $\pm$ 0.020 &    5  \\
2013-05-22.72  &  435.22   &   95.22   &           ---        &  17.002 $\pm$ 0.030 &  15.850 $\pm$ 0.011 &  15.390 $\pm$ 0.012 &  15.048 $\pm$ 0.023 &          ---        &          ---        &          ---        &    1  \\
2013-05-23.73  &  436.23   &   96.23   &           ---        &  17.075 $\pm$ 0.047 &  15.950 $\pm$ 0.022 &  15.456 $\pm$ 0.019 &  15.129 $\pm$ 0.034 &          ---        &          ---        &          ---        &    1  \\
2013-05-24.74  &  437.24   &   97.24   &           ---        &  17.191 $\pm$ 0.062 &  16.017 $\pm$ 0.022 &  15.543 $\pm$ 0.020 &  15.204 $\pm$ 0.037 &          ---        &          ---        &          ---        &    1  \\
2013-05-25.23  &  437.73   &   97.73   &           ---        &  17.383 $\pm$ 0.125 &  16.092 $\pm$ 0.044 &          ---        &          ---        &  16.621 $\pm$ 0.052 &  15.738 $\pm$ 0.047 &  15.605 $\pm$ 0.033 &    5  \\
2013-05-25.83  &  438.33   &   98.33   &           ---        &  17.213 $\pm$ 0.089 &  16.081 $\pm$ 0.032 &  15.613 $\pm$ 0.021 &  15.270 $\pm$ 0.036 &          ---        &          ---        &          ---        &    1  \\
2013-05-26.72  &  439.22   &   99.22   &           ---        &  17.417 $\pm$ 0.040 &  16.191 $\pm$ 0.015 &  15.677 $\pm$ 0.017 &  15.293 $\pm$ 0.033 &          ---        &          ---        &          ---        &    1  \\
2013-05-27.23  &  439.73   &   99.73   &           ---        &  17.671 $\pm$ 0.021 &  16.264 $\pm$ 0.011 &          ---        &          ---        &  16.816 $\pm$ 0.008 &  15.842 $\pm$ 0.010 &  15.749 $\pm$ 0.010 &    5  \\
2013-05-27.70  &  440.20   &  100.20   &   18.856 $\pm$ 0.355 &  17.505 $\pm$ 0.037 &  16.254 $\pm$ 0.014 &  15.729 $\pm$ 0.017 &  15.335 $\pm$ 0.032 &          ---        &          ---        &          ---        &    1  \\
2013-05-28.76  &  441.26   &  101.26   &   18.513 $\pm$ 0.293 &  17.638 $\pm$ 0.038 &  16.382 $\pm$ 0.014 &  15.829 $\pm$ 0.019 &  15.417 $\pm$ 0.033 &          ---        &          ---        &          ---        &    1  \\
2013-05-29.23  &  441.73   &  101.73   &           ---        &  17.841 $\pm$ 0.024 &  16.517 $\pm$ 0.029 &          ---        &          ---        &  17.030 $\pm$ 0.011 &  16.018 $\pm$ 0.008 &  15.922 $\pm$ 0.014 &    5  \\
2013-05-30.23  &  442.73   &  102.73   &           ---        &  18.019 $\pm$ 0.021 &  16.605 $\pm$ 0.014 &          ---        &          ---        &  17.199 $\pm$ 0.014 &  16.128 $\pm$ 0.011 &  16.017 $\pm$ 0.010 &    5  \\
2013-05-30.80  &  443.30   &  103.30   &           ---        &  17.844 $\pm$ 0.038 &  16.594 $\pm$ 0.015 &  16.004 $\pm$ 0.017 &          ---        &          ---        &          ---        &          ---        &    1  \\
2013-05-31.24  &  443.74   &  103.74   &           ---        &  18.170 $\pm$ 0.039 &  16.736 $\pm$ 0.020 &          ---        &          ---        &  17.345 $\pm$ 0.015 &  16.241 $\pm$ 0.013 &          ---        &    5  \\
2013-06-01.23  &  444.73   &  104.73   &           ---        &  18.259 $\pm$ 0.025 &  16.800 $\pm$ 0.018 &          ---        &          ---        &  17.422 $\pm$ 0.012 &  16.289 $\pm$ 0.010 &  16.204 $\pm$ 0.020 &    5  \\
2013-06-02.23  &  445.73   &  105.73   &           ---        &  18.485 $\pm$ 0.051 &  16.894 $\pm$ 0.027 &          ---        &          ---        &  17.548 $\pm$ 0.018 &  16.370 $\pm$ 0.015 &  16.256 $\pm$ 0.023 &    5  \\
2013-06-04.18  &  447.68   &  107.68   &           ---        &  18.599 $\pm$ 0.046 &  17.113 $\pm$ 0.023 &          ---        &          ---        &  17.734 $\pm$ 0.021 &  16.529 $\pm$ 0.017 &  16.411 $\pm$ 0.022 &    5  \\
2013-06-04.78  &  448.28   &  108.28   &           ---        &          ---        &  17.048 $\pm$ 0.020 &  16.399 $\pm$ 0.022 &          ---        &          ---        &          ---        &          ---        &    1  \\
2013-06-07.18  &  450.68   &  110.68   &           ---        &  18.700 $\pm$ 0.047 &  17.207 $\pm$ 0.029 &          ---        &          ---        &  17.888 $\pm$ 0.022 &  16.744 $\pm$ 0.018 &  16.619 $\pm$ 0.026 &    5  \\
2013-06-10.22  &  453.72   &  113.72   &           ---        &  18.806 $\pm$ 0.084 &  17.278 $\pm$ 0.031 &          ---        &          ---        &  17.962 $\pm$ 0.025 &  16.715 $\pm$ 0.026 &  16.691 $\pm$ 0.036 &    5  \\
2013-06-12.22  &  455.72   &  115.72   &           ---        &          ---        &  17.316 $\pm$ 0.028 &          ---        &          ---        &  18.003 $\pm$ 0.024 &  16.794 $\pm$ 0.021 &  16.651 $\pm$ 0.030 &    5  \\
2013-06-12.68  &  456.18   &  116.18   &   19.743 $\pm$ 0.527 &          ---        &  17.291 $\pm$ 0.031 &  16.632 $\pm$ 0.025 &  16.146 $\pm$ 0.037 &          ---        &          ---        &          ---        &    1  \\
2013-06-13.72  &  457.22   &  117.22   &   19.625 $\pm$ 0.411 &          ---        &  17.293 $\pm$ 0.023 &  16.627 $\pm$ 0.024 &  16.161 $\pm$ 0.036 &          ---        &          ---        &          ---        &    1  \\
2013-06-16.13  &  459.63   &  119.63   &           ---        &  18.911 $\pm$ 0.044 &  17.414 $\pm$ 0.018 &          ---        &          ---        &  18.053 $\pm$ 0.042 &  16.806 $\pm$ 0.017 &  16.727 $\pm$ 0.024 &    5  \\
2013-06-17.20  &  460.70   &  120.70   &           ---        &  18.885 $\pm$ 0.038 &  17.362 $\pm$ 0.110 &          ---        &          ---        &          ---        &          ---        &          ---        &    5  \\
2013-06-19.72  &  463.22   &  123.22   &           ---        &          ---        &  17.446 $\pm$ 0.036 &  16.749 $\pm$ 0.026 &  16.292 $\pm$ 0.038 &          ---        &          ---        &          ---        &    1  \\
2013-06-20.10  &  463.60   &  123.60   &           ---        &          ---        &  17.478 $\pm$ 0.026 &          ---        &          ---        &  18.052 $\pm$ 0.020 &          ---        &          ---        &    5  \\
2013-06-21.73  &  465.23   &  125.23   &           ---        &          ---        &  17.463 $\pm$ 0.053 &  16.700 $\pm$ 0.034 &  16.245 $\pm$ 0.045 &          ---        &          ---        &          ---        &    1  \\
2013-06-22.05  &  465.55   &  125.55   &           ---        &  18.790 $\pm$ 0.065 &  17.454 $\pm$ 0.021 &          ---        &          ---        &  18.007 $\pm$ 0.021 &  16.851 $\pm$ 0.012 &  16.794 $\pm$ 0.022 &    5  \\
2013-06-24.05  &  467.55   &  127.55   &           ---        &  19.004 $\pm$ 0.086 &          ---        &          ---        &          ---        &  18.032 $\pm$ 0.018 &  16.901 $\pm$ 0.013 &  16.770 $\pm$ 0.015 &    5  \\
2013-06-26.05  &  469.55   &  129.55   &           ---        &  18.950 $\pm$ 0.048 &  17.519 $\pm$ 0.018 &          ---        &          ---        &  18.091 $\pm$ 0.015 &  16.896 $\pm$ 0.011 &  16.851 $\pm$ 0.015 &    5  \\
2013-07-02.06  &  475.56   &  135.56   &           ---        &  18.963 $\pm$ 0.029 &  17.623 $\pm$ 0.018 &          ---        &          ---        &  18.190 $\pm$ 0.014 &  16.912 $\pm$ 0.012 &  16.898 $\pm$ 0.013 &    5  \\
2013-07-03.06  &  476.56   &  136.56   &           ---        &  18.936 $\pm$ 0.032 &  17.533 $\pm$ 0.021 &          ---        &          ---        &          ---        &  16.880 $\pm$ 0.017 &  16.911 $\pm$ 0.024 &    5  \\
\\ \hline
  UT Date&JD&Phase$^{a}$&$U$&$B$&$V$&$R$&$I$&$g$&$r$&$i$&Tel$^{b}$ \\
  (yyyy/mm/dd)&2456000+&(day)&(mag)&(mag)&(mag)&(mag)&(mag)&(mag)&(mag)&(mag)& /Inst\\
  \hline
2013-07-04.21  &  477.71   &  137.71   &           ---        &  18.933             &          ---        &          ---        &          ---        &          ---        &          ---        &  16.990 $\pm$ 0.022 &    5  \\
2013-07-05.21  &  478.71   &  138.71   &           ---        &  19.027 $\pm$ 0.052 &  17.588 $\pm$ 0.020 &          ---        &          ---        &  18.178 $\pm$ 0.024 &  16.956 $\pm$ 0.011 &  16.937 $\pm$ 0.014 &    5  \\
2013-07-06.21  &  479.71   &  139.71   &           ---        &  18.875 $\pm$ 0.052 &  17.579 $\pm$ 0.026 &          ---        &          ---        &  18.168 $\pm$ 0.020 &  16.983 $\pm$ 0.015 &  16.965 $\pm$ 0.021 &    5  \\
2013-07-07.19  &  480.69   &  140.69   &           ---        &  18.936 $\pm$ 0.068 &  17.624 $\pm$ 0.035 &          ---        &          ---        &  18.188 $\pm$ 0.022 &  16.952 $\pm$ 0.020 &  16.952 $\pm$ 0.017 &    5  \\
2013-07-11.15  &  484.65   &  144.65   &           ---        &  18.865 $\pm$ 0.050 &  17.611 $\pm$ 0.024 &          ---        &          ---        &  18.185 $\pm$ 0.019 &  16.968 $\pm$ 0.017 &  17.067 $\pm$ 0.020 &    5  \\
2013-07-13.15  &  486.65   &  146.65   &           ---        &  18.910 $\pm$ 0.050 &  17.604 $\pm$ 0.028 &          ---        &          ---        &  18.206 $\pm$ 0.016 &  16.993 $\pm$ 0.014 &  17.026 $\pm$ 0.019 &    5  \\
2013-07-15.18  &  488.68   &  148.68   &           ---        &          ---        &  17.773 $\pm$ 0.066 &          ---        &          ---        &  18.258 $\pm$ 0.078 &  17.089 $\pm$ 0.045 &          ---        &    5  \\
2013-07-21.18  &  494.68   &  154.68   &           ---        &  19.072 $\pm$ 0.111 &  17.914 $\pm$ 0.043 &          ---        &          ---        &  18.240 $\pm$ 0.034 &  17.047 $\pm$ 0.021 &  17.215 $\pm$ 0.024 &    5  \\
2013-07-26.18  &  499.68   &  159.68   &           ---        &  19.058 $\pm$ 0.053 &  17.814 $\pm$ 0.035 &          ---        &          ---        &  18.260 $\pm$ 0.032 &  17.073 $\pm$ 0.018 &  17.197 $\pm$ 0.029 &    5  \\
2013-07-30.72  &  504.22   &  164.22   &           ---        &  19.044 $\pm$ 0.038 &  17.753 $\pm$ 0.021 &          ---        &          ---        &          ---        &          ---        &          ---        &    5  \\
2013-07-31.43  &  504.93   &  164.93   &           ---        &  19.090 $\pm$ 0.111 &  17.708 $\pm$ 0.044 &          ---        &          ---        &          ---        &  17.096 $\pm$ 0.027 &  17.257 $\pm$ 0.046 &    3  \\
2013-08-01.17  &  505.67   &  165.67   &           ---        &  19.070 $\pm$ 0.150 &          ---        &          ---        &          ---        &  18.384 $\pm$ 0.030  &          ---        &          ---        &    5  \\
2013-08-04.12  &  508.62   &  168.62   &           ---        &  19.039 $\pm$ 0.113 &  17.900 $\pm$ 0.044 &          ---        &          ---        &  18.350 $\pm$ 0.025 &  17.129 $\pm$ 0.017 &  17.285 $\pm$ 0.025 &    5  \\
2013-08-06.18  &  510.68   &  170.68   &           ---        &  19.047 $\pm$ 0.068 &  17.886 $\pm$ 0.031 &          ---        &          ---        &  18.392 $\pm$ 0.024 &  17.247 $\pm$ 0.037 &  17.413 $\pm$ 0.035 &    5  \\
2013-08-11.97  &  516.47   &  176.47   &           ---        &  19.073 $\pm$ 0.125 &  18.005 $\pm$ 0.049 &          ---        &          ---        &          ---        &  17.186 $\pm$ 0.022 &  17.385 $\pm$ 0.036 &    5  \\
2013-08-14.01  &  518.51   &  178.51   &           ---        &  19.074 $\pm$ 0.080 &  18.035 $\pm$ 0.036 &          ---        &          ---        &  18.487 $\pm$ 0.028 &  17.231 $\pm$ 0.017 &  17.422 $\pm$ 0.035 &    5  \\
2013-08-16.99  &  521.49   &  181.49   &           ---        &          ---        &  18.099 $\pm$ 0.051 &          ---        &          ---        &  18.563 $\pm$ 0.036 &          ---        &  17.428 $\pm$ 0.035 &    5  \\
2013-08-20.15  &  524.65   &  184.65   &           ---        &          ---        &          ---        &          ---        &          ---        &  18.613 $\pm$ 0.125 &  17.227 $\pm$ 0.041 &  17.433 $\pm$ 0.072 &    5  \\
2013-08-20.42  &  524.92   &  184.92   &           ---        &  19.089 $\pm$ 0.122 &  18.078 $\pm$ 0.051 &          ---        &          ---        &          ---        &          ---        &          ---        &    5  \\
2013-08-24.38  &  528.88   &  188.88   &           ---        &          ---        &  18.002 $\pm$ 0.060 &          ---        &          ---        &          ---        &          ---        &          ---        &    3  \\
2013-08-25.36  &  529.86   &  189.86   &           ---        &          ---        &          ---        &          ---        &          ---        &          ---        &  17.267 $\pm$ 0.036 &  17.526 $\pm$ 0.071 &    3  \\
    \hline
  \end{longtable}
  \begin{longtable}
  {c c r c c c c c c c l}
  \caption*{$Swift$~UVOT photometry}\\
  \hline
    UT Date&JD&Phase$^{a}$     & $uvw1$& $uvw2$& $uvm2$& $uvu$& $uvb$& $uvv$&Tel$^{b}$ & \\
    (yyyy/mm/dd)&2456000+&(day)&(mag)& (mag)& (mag)& (mag)& (mag)& (mag)& /Inst & \\ \hline
2013-02-21.18   &  344.68   &    4.68   &   13.404 $\pm$ 0.038 &  13.434 $\pm$ 0.042 &  13.301 $\pm$ 0.036 &          ---        &          ---        &          ---        &    4  \\
2013-02-21.94   &  345.44   &    5.44   &   13.486 $\pm$ 0.039 &  13.589 $\pm$ 0.052 &  13.442 $\pm$ 0.047 &  13.670 $\pm$ 0.052 &  15.010 $\pm$ 0.049 &  15.094 $\pm$ 0.056 &    4  \\
2013-02-22.08   &  345.58   &    5.58   &   13.489 $\pm$ 0.037 &  13.629 $\pm$ 0.044 &          ---        &          ---        &          ---        &          ---        &    4  \\
2013-02-22.51   &  346.01   &    6.01   &   13.616 $\pm$ 0.056 &  13.747 $\pm$ 0.071 &          ---        &          ---        &          ---        &          ---        &    4  \\
2013-02-24.04   &  347.54   &    7.54   &   13.734 $\pm$ 0.035 &  14.133 $\pm$ 0.041 &  13.830 $\pm$ 0.035 &          ---        &          ---        &          ---        &    4  \\
2013-02-24.90   &  348.40   &    8.40   &   13.848 $\pm$ 0.043 &  14.363 $\pm$ 0.052 &  14.007 $\pm$ 0.037 &  13.819 $\pm$ 0.050 &  15.010 $\pm$ 0.046 &  14.962 $\pm$ 0.047 &    4  \\
2013-02-24.98   &  348.48   &    8.48   &           ---        &          ---        &  14.083 $\pm$ 0.059 &          ---        &          ---        &          ---        &    4  \\
2013-02-27.30   &  350.80   &   10.80   &           ---        &  14.895 $\pm$ 0.072 &          ---        &          ---        &          ---        &          ---        &    4  \\
2013-02-27.98   &  351.48   &   11.48   &   14.395 $\pm$ 0.058 &  14.992 $\pm$ 0.072 &  14.742 $\pm$ 0.060 &  13.997 $\pm$ 0.051 &  15.072 $\pm$ 0.047 &  15.018 $\pm$ 0.049 &    4  \\
2013-03-03.09   &  354.59   &   14.59   &   15.084 $\pm$ 0.060 &  15.792 $\pm$ 0.075 &  15.662 $\pm$ 0.063 &  14.340 $\pm$ 0.052 &  15.237 $\pm$ 0.047 &  15.132 $\pm$ 0.050 &    4  \\
2013-03-06.13   &  357.63   &   17.63   &   15.635 $\pm$ 0.063 &  16.577 $\pm$ 0.082 &  16.574 $\pm$ 0.072 &  14.608 $\pm$ 0.052 &  15.258 $\pm$ 0.047 &  15.165 $\pm$ 0.049 &    4  \\
2013-03-14.17   &  365.67   &   25.67   &   17.408 $\pm$ 0.116 &  18.759 $\pm$ 0.261 &  18.859 $\pm$ 0.320 &  15.680 $\pm$ 0.068 &  15.552 $\pm$ 0.054 &  15.220 $\pm$ 0.061 &    4  \\
2013-03-18.21   &  369.71   &   29.71   &   17.710 $\pm$ 0.127 &  19.461 $\pm$ 0.364 &  19.231 $\pm$ 0.335 &  16.203 $\pm$ 0.066 &  15.739 $\pm$ 0.049 &  15.303 $\pm$ 0.051 &    4  \\
2013-03-20.97   &  372.47   &   32.47   &   18.073 $\pm$ 0.153 &  19.120 $\pm$ 0.264 &  19.430 $\pm$ 0.385 &  16.458 $\pm$ 0.068 &  15.896 $\pm$ 0.049 &  15.354 $\pm$ 0.049 &    4  \\
2013-03-27.97   &  379.47   &   39.47   &   18.913 $\pm$ 0.299 &  19.516 $\pm$ 0.549 &          ---        &  17.096 $\pm$ 0.095 &  16.161 $\pm$ 0.054 &          ---        &    4  \\
2013-04-01.39   &  383.89   &   43.89   &   18.754 $\pm$ 0.266 &  19.548 $\pm$ 0.390 &  19.785 $\pm$ 0.538 &  17.276 $\pm$ 0.101 &  16.195 $\pm$ 0.053 &  15.439 $\pm$ 0.053 &    4  \\
2013-04-05.20   &  387.70   &   47.70   &   18.935 $\pm$ 0.294 &  21.232 $\pm$ 1.634 &  19.854 $\pm$ 0.556 &  17.431 $\pm$ 0.105 &  16.391 $\pm$ 0.054 &  15.565 $\pm$ 0.053 &    4  \\
2013-04-10.84   &  393.34   &   53.34   &   19.158 $\pm$ 0.359 &  20.750 $\pm$ 1.075 &  20.042 $\pm$ 0.671 &  17.637 $\pm$ 0.120 &  16.472 $\pm$ 0.055 &  15.548 $\pm$ 0.053 &    4  \\
2013-04-21.92   &  404.42   &   64.42   &   19.450 $\pm$ 0.628 &          ---        &          ---        &          ---        &          ---        &          ---        &    4  \\
2013-04-26.01   &  408.51   &   68.51   &   19.505 $\pm$ 0.487 &          ---        &  20.392 $\pm$ 1.262 &  17.890 $\pm$ 0.167 &  16.646 $\pm$ 0.087 &  15.615 $\pm$ 0.086 &    4  \\
2013-05-01.32   &  413.82   &   73.82   &   20.646 $\pm$ 1.300 &  20.560 $\pm$ 0.882 &  20.642 $\pm$ 1.126 &  18.092 $\pm$ 0.160 &  16.728 $\pm$ 0.058 &  15.632 $\pm$ 0.054 &    4  \\
2013-05-14.26   &  426.76   &   86.76   &   21.116 $\pm$ 2.278 &  20.650 $\pm$ 1.277 &          ---        &  18.971 $\pm$ 0.456 &  16.825 $\pm$ 0.080 &  15.821 $\pm$ 0.080 &    4  \\
2013-05-16.00   &  429.50   &   89.50   &           ---        &          ---        &  20.689 $\pm$ 1.217 &  18.496 $\pm$ 0.233 &  16.925 $\pm$ 0.066 &  15.858 $\pm$ 0.062 &    4  \\
2013-05-20.95   &  433.45   &   93.45   &   20.929 $\pm$ 1.738 &  21.652 $\pm$ 2.536 &  20.850 $\pm$ 1.421 &  18.708 $\pm$ 0.279 &  17.223 $\pm$ 0.074 &  16.050 $\pm$ 0.067 &    4  \\
2013-05-30.95   &  443.45   &  103.45   &   20.721 $\pm$ 1.389 &  21.184 $\pm$ 1.547 &  21.605 $\pm$ 2.708 &  19.429 $\pm$ 0.464 &  18.155 $\pm$ 0.115 &  16.923 $\pm$ 0.096 &    4  \\
  \hline
  \end{longtable}
\begin{flushleft}
  $^{a}$ with reference to the explosion epoch JD 2456340.0\\
  $^{b}$ 1: 104-cm Sampurnanand Telescope (ST) and 130-cm Devasthal fast optical telescope (DFOT), ARIES, India; 2: Faulkes Telescope South; 3: Faulkes Telescope North; 4:~Swift UVOT data; 5:~1-m class telescopes of Las Cumbres Observatory Global Telescope (LCOGT) network; 6: IAU circular report.\\
\end{flushleft}
\end{landscape}
\twocolumn

\section{Spectroscopy}
\label{app:spectroscopy}

 Spectroscopic data were reduced in the \iraf\ environment. All frames are corrected by bias subtraction and flat fielding. Cosmic ray rejection on each frame was done using
 Laplacian kernel detection algorithm for spectra \citep[L.A.Cosmic;][]{2001PASP..113.1420V}.
 One-dimensional spectra were extracted using {\it apall} task which is based on
 optimal extraction algorithm by \citet{1986PASP...98..609H}.
 Wavelength calibration was performed with the help of the {\it identify} task by using the arc spectra obtained for all telescopes during each night.
 The position of \ion{O}{I} emission skyline at 5577\AA\, was used to check the wavelength
 calibration and deviations were found to lie between 0.3 to 5.5\AA, and this was corrected by
 applying a linear shift in wavelength.

 The flux calibration of wavelength-calibrated spectra was done using \textit{standard, sensfunc}
 and \textit{calibrate} tasks. We used
 spectrophotometric standard fluxes from \citet{1990AJ.....99.1621O,1994PASP..106..566H}.
 All  spectra were tied to an absolute flux scale using zeropoints determined from \ubvri\ magnitudes.
 To tie the spectra with photometry, individual spectrum was multiplied by wavelength
 dependent polynomial function and its $ BVRI $ filter-response convolved fluxes were compared
 with photometric fluxes at corresponding epoch. The multiplied polynomial was tuned to minimize
 the flux difference and obtain the tied spectrum.
 The one-dimensional spectra were corrected for
 heliocentric velocity of the
 host galaxy (1374 \kms; see \S\ref{sec:intro}) using {\it dopcor} tasks.

\FloatBarrier
\begin{table*}
  \caption{Journal of spectroscopic observations of \sn.
  The spectral observations are made at 25 phases from 2 to 184d.}
  \label{tab:speclog}
  \begin{tabular}{lc cc cc}
    \hline
    UT Date       &JD      &Phase$^{a}$&Telescope$^{c}$&    Range$^{b}$& Exposure\\
    (yyyy-mm-dd.dd) &2456000+&(days)     &               & \mum          & (s)     \\ \hline

2013-02-18.70  &  342.20  &   2.20 &  FTS  &  0.32-1.00  &   1500   \\
2013-02-19.74  &  343.24  &   3.24 &  FTS  &  0.32-1.00  &   1500   \\
2013-02-24.71  &  348.21  &   8.21 &  FTS  &  0.32-1.00  &   1500   \\
2013-02-28.59  &  352.09  &  12.09 &  FTN  &  0.32-1.00  &   1800   \\
2013-03-01.08  &  352.58  &  12.58 &  PEN  &  0.00-0.00  &   1800   \\
2013-03-01.50  &  353.00  &  13.00 &  FTN  &  0.32-1.00  &   1800   \\
2013-03-02.51  &  354.01  &  14.01 &  FTN  &  0.32-1.00  &   1800   \\
2013-03-03.51  &  355.01  &  15.01 &  FTN  &  0.32-1.00  &   1800   \\
2013-03-04.59  &  356.09  &  16.09 &  FTN  &  0.32-1.00  &   1800   \\
2013-03-06.71  &  358.21  &  18.21 &  FTS  &  0.32-1.00  &   2000   \\
2013-03-17.93  &  369.43  &  29.43 &  HCT  &  0.38-0.84  &   1800   \\
2013-03-19.63  &  371.13  &  31.13 &  FTS  &  0.32-1.00  &   2000   \\
2013-03-31.58  &  383.08  &  43.08 &  FTS  &  0.32-1.00  &   2400   \\
2013-05-01.84  &  414.34  &  74.34 &  HCT  &  0.38-0.84  &   1200   \\
2013-05-03.35  &  415.85  &  75.85 &  FTN  &  0.32-1.00  &   3600   \\
2013-05-04.33  &  416.83  &  76.83 &  FTN  &  0.32-1.00  &   3600   \\
2013-05-04.85  &  417.35  &  77.35 &  HCT  &  0.38-0.68  &   1800   \\
2013-05-05.08  &  417.58  &  77.58 &  PEN  &  0.00-0.00  &   1200   \\
2013-05-09.57  &  422.07  &  82.07 &  FTN  &  0.32-1.00  &   3600   \\
2013-05-15.85  &  428.35  &  88.35 &  HCT  &  0.38-0.84  &   1200   \\
2013-06-14.64  &  458.14  & 118.14 &  HCT  &  0.38-0.68  &   2400   \\
2013-07-13.34  &  486.84  & 146.84 &  FTN  &  0.32-1.00  &   3600   \\
2013-08-09.25  &  513.75  & 173.75 &  FTN  &  0.32-1.00  &   3600   \\
2013-08-18.25  &  522.75  & 182.75 &  FTN  &  0.32-1.00  &   3600   \\
2013-08-19.25  &  523.75  & 183.75 &  FTN  &  0.32-1.00  &   3600  	\\
    \hline
  \end{tabular}
\begin{flushleft}
  $^{a}$ With reference to the explosion time JD 2456340.0\\
  $^{b}$ For transmission $\ge$50\%\\
  $^{c}$ HCT : HFOSC on 2 m Himalyan Chandra Telescope, India; FTN : FLOYDS on 2 m Faulkes Telescope
North, Hawaii; FTS: FLOYDS on 2 m Faulkes Telescope South, Australia; PEN: B\&C spectrograph on 1.22 m Galileo Telescope, Italy. \\
\end{flushleft}
\end{table*}
\FloatBarrier

\end{document}